\RequirePackage{fix-cm} 
\RequirePackage{amsmath}

\documentclass[10pt,journal,twocolumn]{IEEEtran} 
\usepackage{blindtext}
\usepackage{graphicx}
\usepackage{graphics}
\usepackage{url}
\usepackage{verbatim}
\usepackage{balance}
\usepackage[scientific-notation=true]{siunitx}
\usepackage{todonotes}
\usepackage{units}
\usepackage{color}
\usepackage{textcomp}
\usepackage{todonotes}

\newcommand{\accept}[2]{#2}

\usepackage{soul}
\usepackage{url}
\usepackage{hyperref}
\hypersetup{
    colorlinks=true,
    linkcolor=blue,
    filecolor=magenta,      
    urlcolor=cyan,
}

\graphicspath{{./figures/}}

\begin{document}
\title{{\it Pound}: A ROS node for Reducing Delay and Jitter in Wireless Multi-Robot Networks}

\author{Danilo Tardioli, \and Ramviyas Parasuraman, and \and Petter \"{O}gren
    
\thanks{D. Tardioli is with Centro Universitario de la Defensa, Zaragoza, 50090 Spain. email: {\it dantard@unizar.es}. R. Parasuraman is with Purdue University, West Lafayette 47906, USA. email: {\it ramviyas@purdue.edu}. P. \"{O}gren is with KTH Royal Institute of Technology, Stockholm 10044, Sweden. email: { \it petter@kth.se}. } }

\maketitle
\begin{abstract}
The Robot Operating System (ROS) is rapidly becoming the \textit{de facto} framework for building robotics systems, thanks to its flexibility and the large acceptance that it has received in the robotics community. With the growth of its popularity, it has started to be used in multi-robot systems as well. However, the TCP connections that the platform relies on for connecting the so-called ROS nodes, presents several issues in terms of limited-bandwidth, delays and jitter, when used in wireless ad-hoc networks.

In this paper, we present a thorough analysis of the problem, and propose a new ROS node called \textit{Pound} to improve the wireless communication performance. 
\textit{Pound} allows the use of multiple ROS cores and introduces a priority scheme, which allows favoring more important flows over less important ones, reducing delay and jitter over single-hop and  multi-hop networks. 
We compare \textit{Pound} to the state of the art solutions, and show that it performs equally well, or better in all the test cases, including a control-over-network example.

\end{abstract}
%

\section{Introduction}
\IEEEPARstart{R}{obot} Operating System (ROS)\footnote{ \url{http://wiki.ros.org}}, introduced in 2007, is today one of the most used software development framework in the robotics community. 
ROS is a open source robotics middleware and a flexible framework for writing robot software made by a collection of tools, libraries, and conventions that aim to simplify the task of creating complex and robust robot behavior, across a wide variety of robotic platforms \cite{ROS}.
It was created with the goal of encouraging collaborative robot software development, in which different problems can be solved by different groups,
allowing the creation of more complex combined systems. In fact, \cite{Chitic2014} presents an overview of various available robotics middlewares and shows how ROS stands out as the most viable middleware especially in the context of multi-robot systems. The objective of ROS is mainly to conduct rapid research and development in both academia and industries. For instance, industrial robots makers such as ABB and KUKA\cite{bischoff2011kuka} are exploring the possibility of offering a ROS interface on their products. 

ROS systems are based on so-called \textit{nodes}, each of them in charge of performing a specific task and offering the results to the user (for example one node can be capable of interfacing with a hardware LIDAR sensor and output the range readings). The output of each node can be used directly by the user or by other nodes that use that information as its input (for example a Simultaneous Localization And Mapping (SLAM) node can use the LIDAR readings to build a map of the environment).
These nodes can have multiple inputs and outputs, and
the exchange of information takes place through the so-called \textit{topics}, named ports that \textit{hide}  simple network connections.

When the different nodes reside in the same computer these connection are local ---they use the \textit{loopback} interface---,
and when they reside on different computers in the same robot they are usually connected through a wired IP network.
In both cases and unless the nodes generate an extraordinary high amount of data, neither the communication bandwidth nor the stability of the connection presents a problem, especially in modern systems equipped with Gigabit Ethernet cards.

\begin{figure}[t]
	  \centering
	  \includegraphics[width=0.98\columnwidth]{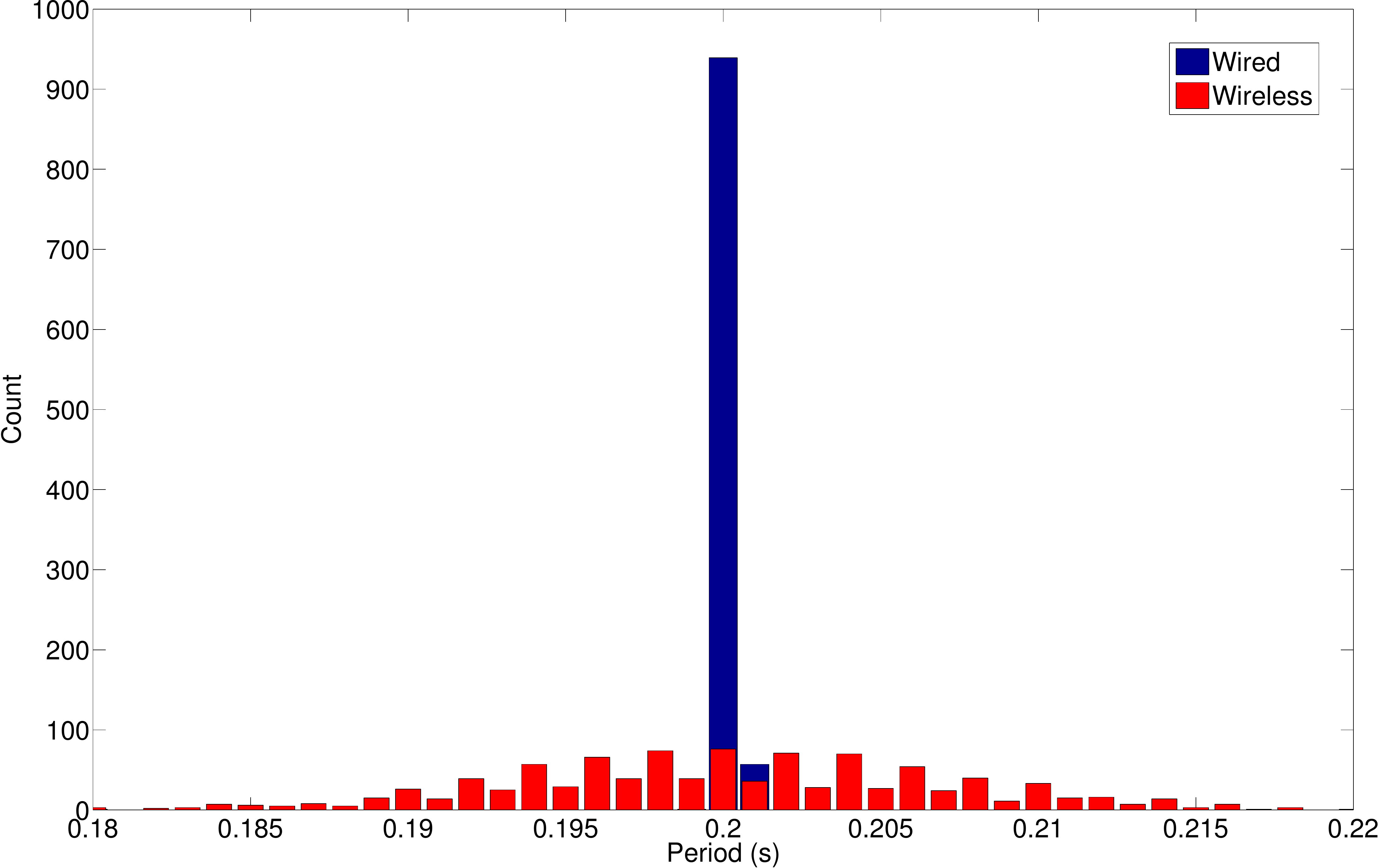}
	  \caption{Comparison between wired and wireless configuration. Period measured at the destination node in a system in which a flow with period of 200 ms is established over wireless and wired connections.}
	  \label{fig:wired_vs_wireless}
 \end{figure}

\accept{However,}{ROS was originally created to distribute tasks, algorithms and also computational burden between different units to promote specialization and collaboration at different levels. However, this distribution has been historically limited to a single machine or, sometimes, to different machines installed in the same physical robot. Even if it is possible to develop multi robot applications using ROS, they have not been actively supported in terms of adapting the framework in any sense. The successful execution of multi-robot missions,
is therefore often dependent on meeting a set of requirements from a communication point of view.
The timing of the data delivery, jitter or  excessive delay can have significant impact on the control loop and, with it, the results.
Even if ROS behaves reasonably well when dealing with local or wired communication,
the situation significantly changes when the nodes that need to exchange information reside in computers that are connected via a wireless network.
In this case, the reliability of the communication is a couple of orders of magnitude lower, the available bandwidth is much more limited and  retransmissions increase the jitter significantly, as can be seen in Figure~\ref{fig:wired_vs_wireless}}.

\subsection{Relation to Multi-Robot applications}

\accept{}{As explained above, the undesired effects of wireless communication are profound when the timing of the data is fundamental for the goal of the task. This is quite frequent in distributed applications where the exchanged data have a periodic and perishable nature. This is the case, for example, in the collaborative or distributed construction of a map (see \cite{7440564}, \cite{tardioli2016robot}, \cite{simmons2000coordination} or \cite{tardioli2014proof} where the laser  and localization data that reach the base station must do so coherently and within firm time windows to allow the task to be completed successfully) or cooperative localization (see \cite{tardioli2016robot} where LIDAR data sent to the base station is used to feed a particle filter) or in audio and video streaming over multi-hop networks as done in \cite{sicignano2011real}. Note that these primary flows are not usually the only ones present in the network; they must share the available bandwidth with less important flows without being perturbed, which creates a need for prioritizing the different flows.
Further, the situation gets worse when the communication is involved in a perception-actuation loop with or without a man in the middle (formation control\cite{Lima201568}, teleoperation, distributed control \cite{6548083}, etc.). In this case, not only is the communication  bidirectional which means that the prioritization of the traffic must be network-wide but the deadlines are also strict rather than firm; the late arrival of data can cause serious issues such as damages and jeopardise the task completion.
}



\accept{In this paper, we propose \textit{Pound} to address these problems.}{In order to address these problems, we propose an alternative solution for multi-robot communication based on ROS through the development of a new ROS node called \textit{Pound}.}

\textit{Pound} \accept{combines}{behaves as a kind of proxy, combining} the use of multiple ROS cores with a priority scheme, allowing the user to assign levels of importance to different topics with the end of reducing the jitter and the delay for data that are considered more important. We open source the \textit{Pound} for the robotics community to use and develop further. It is available in {\tt github}\footnote{\url{https://github.com/dantard/unizar-pound-ros-pkg/}}.

To investigate the performance of \textit{Pound}, we compare it to the state-of-the-art approaches 
such as
\textit{Nimbro}\footnote{\url{https://github.com/AIS-Bonn/nimbro_network}} \cite{Nimbro}, \textit{RT-WMP}\footnote{\url{https://github.com/dantard/unizar-rt-wmp-ros-pkg/}} \cite{RT-WMP},
and the standard ROS communication protocols.
The comparison is made using a scenario with 
 two flows of information of different importance in terms of jitter, delay and bandwidth. We also investigate the effects on a control loop running over a real wireless connection.

The main contributions of this paper are 1) the detailed problem analysis, and 2) the \accept{new proposed ROS node}{proposed solution}. The problem analysis aims to show the behavior of different connection methods especially those provided by the ROS platform \accept{}{but also related approaches already available in the literature} demonstrating that the use \accept{}{of wireless connections, especially in presence} of competing flows and/or scenarios in which multiple hop communication is required, \accept{in wireless environments}{} must be carefully planned.
The ROS node \textit{Pound} goes beyond the state-of-the-art by providing  a system of priority in a per-flow base which allows to favour most important ones while avoiding network congestion and blockages. The theoretical advantages of \textit{Pound} are verified in a thorough evaluation.

The paper is organised as follows. In the next section the related works are presented. In section \ref{sec:problem} the ROS is introduced and the issues related with using it in wireless networks are reported.
In section \ref{sec:solutions_analysed} we present the different solutions analysed in this work and introduce the \textit{Pound} in Section~\ref{sec:Pound}. Then, in section \ref{sec:experiments} we relate the experiments performed and their results. We close the paper with a discussion in section \ref{sec:discussion} and conclusions in section \ref{sec:conclusions}.

\section{Related work}
\label{sec:related_work}
In this section we review the literature on wireless communication in robotic applications.
This problem has been addressed in both the robotics community, and the communications community, with each of them having slightly different problem formulations and corresponding solutions. We begin with the robotics and then discuss the communication community.

From a robotics perspective, it was noted in \cite{Steinfeld2006}, that network parameters such as delay, jitter and  bandwidth (throughput) are the most common influencing metrics in Human-Robot Interaction (HRI) problems. These are important metrics because they affect the robot missions effectiveness and efficiency directly. For instance, \cite{Owen-Hill2014,gergle2006} discuss how the delay between robot camera view and the teleoperator for visual feedback can affect the task performance. Both these works conclude that the smaller the delay is, the faster is the task completion time. Faster completion time is crucial in tasks such as search and rescue during disasters because every second may be utilized in saving more lives in such situations. Moreover, \cite{Haller2015} has demonstrated the importance of jitter as an indicator of the communication performances in a wireless networked robotic system. It also shows how variations in jitter compromises teleoperation performance and proposed a control algorithm to handle the variable communication delay in a WLAN for bilateral teleoperation. An FPGA controller was developed to handle the latency problem in  wireless robot motion control \cite{lomonova2010advanced}. However, such solutions do not address the problem from a communication perspective, but rather from a robotic control theory and  are thus very application specific. 

Birk et. al \cite{birk2009networking} focus on wireless network for teleoperation application especially for search and rescue robots and proposed the Jacobs Intelligent Robotics Library (JIRlib) that features compression, prioritization, and serialization framework to exchange data between robots and control station over typical TCP/IP framework. Although an attractive robotic communication framework, it is not integrated in ROS and hence cannot be readily used. In some works, the problems of wireless communications are dealt with differently. Assuming poor connection quality with standard wireless networks, authors in \cite{owen2013haptic,caccamo2015extending} proposed enhanced visual and haptic feedback mechanisms to help the robot operator perceive directional wireless connectivity and thereby reduce the risk of communication failure. Wireless tethering and spatial diversity techniques are exploited in \cite{parasuraman2013spatial,min2013design,tardioli2015robot}. In \cite{robinson2013aerial,chirwa2014performance}, a Mobile Ad-hoc Network (MANET) using Unmanned Aerial Vehicles (UAV) is proposed for improved communication between robots. 

From a pure communications perspective, the problem with wireless networks is addressed mostly with modifications or adaptations of protocols \cite{tian2005tcp}. For instance, \cite{tian2005tcp} presents the problems with the predominately used TCP/IP protocol in lossy wireless networks and suggest some possible proactive schemes such as TCP-Westwood and TCP-Jersey as replacements (or enhancements) to standard TCP to counter the issue of link instability and high Bit Error Rate (BER). However such enhanced protocols adds overheads which may have a significant negative impact in dynamic robotic networks. Deek et. al \cite{deek2014intelligent} advised the use of channel bonding in typical 802.11n WLAN to improve the data rates. Paasch et. al \cite{Paasch2012} proposed the Multipath TCP (MPTCP) protocol which uses all available interfaces in a terminal to increase the application throughput. However, most of these solutions do not solve the inherent latency and jitter problems in a robot network. Other solutions like \cite{patti2015} aim to provide support traffic scheduling, implementing multiple traffic classes with different priorities transmitted according to a fixed schedule. However, most of these schemes rely on a centralized approach which is not applicable to mobile robots.

Finally, standardizing wireless communication for mobile robots is advocated in \cite{knoll2012wireless, schioler2012wireless} as key area of improvement. In fact, \cite{schioler2012wireless} suggested an architectural framework for mobile robot use cases and applications considering various layers in the OSI reference model. This paper provides one step in that direction, by identifying drawbacks in, and proposing solutions for, the most widely used ROS framework. 

\accept{}{Authors in \cite{Blasco2012} provided a nice comparison of various multi-agent and robotics middleware frameworks, and emphasized the inability of ROS to provide real-time capabilities due to its dependency on TCP/IP.} Only a very limited amount of work can be found in the literature that discusses how ROS handles data flows and proposes solutions to improve the use of ROS in difficult network situations. For instance, \cite{Forouher2014} presents a reporting framework for providing network statistics, such as  bandwidth and frequency, on every ROS nodes. They also further extend the framework using existing ROS tools (diagnostics tool, \textit{rqt\_graph}, etc.) and demonstrate their ideas in applications such as robot health monitoring. In \cite{ARNI}, a similar approach is taken to provide advanced features to monitor and introspect a ROS system through a tool called ARNI\footnote{\url{http://wiki.ros.org/arni}}. However, it's worth noting that these network monitoring tools does not provide new means of communication.

In this paper we go beyond the work described above,  and analyse the ROS framework and existing ROS-compatible network protocol extensions to expose their problems in the context of wireless networked mobile robots. Then, based on our analysis, we propose viable alternatives and solutions depending on the nature of the problems. 

\section{Problem Analysis}
\label{sec:problem}
In this section, we will first give an overview of ROS, and then describe the 
different drawbacks associated with running ROS over a wireless network.
Finally, we investigate the different solutions proposed so far to address the problems.

\subsection{ROS Overview}
\label{sec:ros_overview}

As described above, ROS is a flexible framework for writing robot software. It acts as a meta-operating system for robots as it provides hardware abstraction, low-level device control, inter-processes message-passing and package management. It also provides tools and libraries for building, writing, and running code in a single or across multiple computers. A robotic system can be built as a set of  so-called ROS nodes that, in turn, contain the processes in charge or producing and/or consuming information. ROS nodes are connected through ROS topics, labeled data streams used to pass information among them. 

Beside the publish/subscriber, ROS also supports client/server paradigm through the ROS services.
The exchange of information among nodes is carried out through simple network sockets using both connection-oriented and connectionless protocols called  TCPROS and UDPROS (defined on top of the standard TCP and UDP protocols). Also, it exist the option of disabling the TCP's Nagle's algorithm using the so-called TCP\_Nodelay option. All this, allows having multiple local nodes and/or to distribute them in different computers. A special node called ROS master (or \textit{roscore}) is in charge of establishing the required connections among the nodes. It acts in a similar way as a DNS: when a node is started, it registers itself and its topics (i.e. their address and ports) in the ROS master. When another node needs to subscribe to a specific topic, it asks the master ---by providing the name--- at what address and port that topic is located, and afterwards establish a direct connection with the publisher node.

\begin{figure}[tpb]
\begin{center}
\includegraphics[page=1,width=1.05\columnwidth]{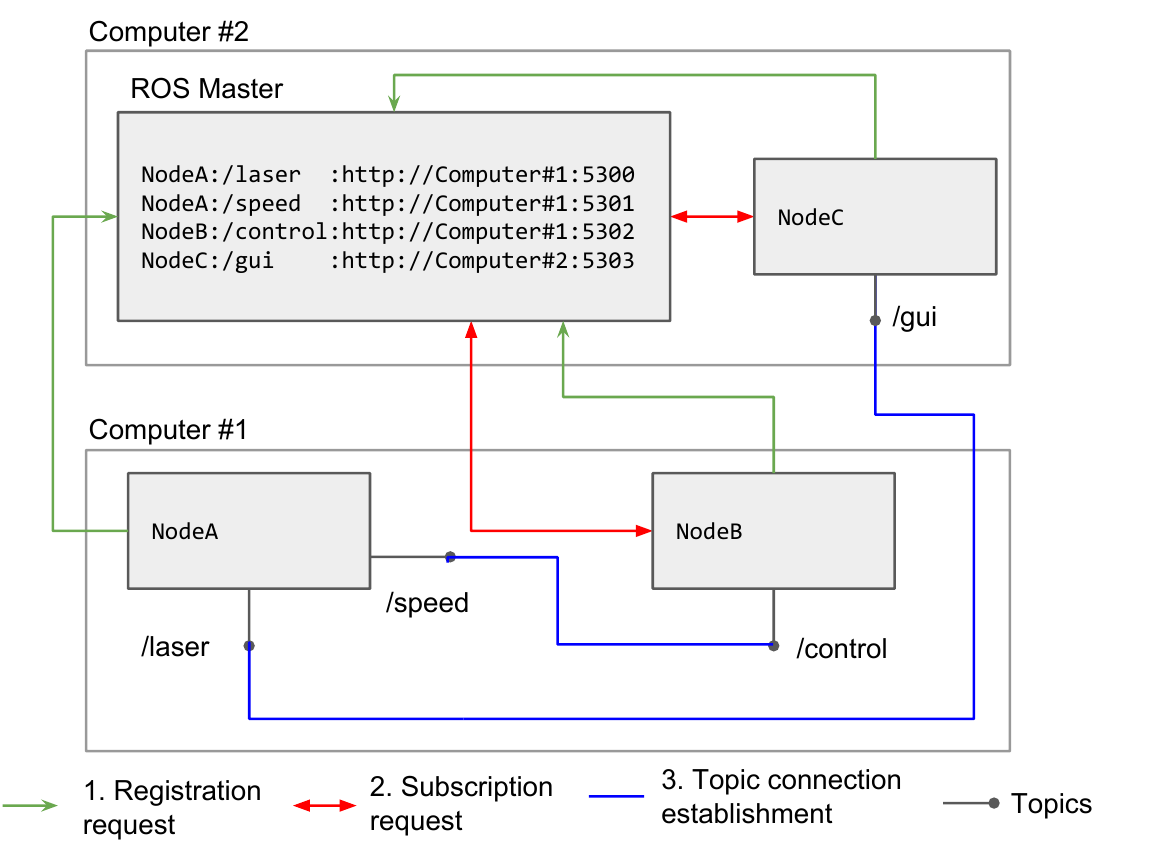}
\end{center}
\caption{\label{fig:ros}\small An overview of a ROS based robotics system. The ROS nodes, once launched, register themselves in the ROS master (which resides in one of the computers), specifying the address and port of their published topics (green lines). When a node wants to subscribe to other node's topic, it asks the master (\textit{roscore}) which responds with the correspondent address and port (red lines). Finally a direct topic connection is established between the publisher and the subscriber (blue lines).}
\end{figure}

In Figure~\ref{fig:ros}, a generic overview of a typical ROS robotic system is presented in which multiple ROS nodes are running (in \textit{Computer \#1}) that publishes/subscribes to ROS topics from another computer (\textit{Computer \#2}) via the ROS master (core) situated on \textit{Computer \#2} as well. Among many references available in the literature for ROS, \cite{ROS} provides a nice overview of how ROS works and how to use it in practical robotic applications.


\paragraph{ROS2: } A second version of ROS (called ROS2) is under development by the ROS community. The main motivations for the ROS2 design are to meet real-time capability requirements and enhance multi-robot communication performance. ROS2 will use the Data Distribution Service (DDS) as its networking middleware because of several advantages such as the distributed nature and Quality of Service (QoS) based configurations. 
However, as ROS2 is still under initial development \accept{}{ 
it is still unclear when a switch from ROS1 to ROS2 will be possible.
Thus,}{} we focus 
this paper on  problems, and corresponding solutions, of the current ROS framework. 

\accept{}{Note that the below mentioned problems, although applicable to ROS as it depend heavily on TCP/IP standard, are generic to any wireless multi-agent systems. The inherent latencies caused by the ROS framework are negligible compared to the network latencies.}

\subsection{Reliability problems}
\label{sec:reliability}
ROS was originally designed for systems in which multiple computers residing  in a single robot were connected through a LAN as a way to distribute the computational load in different units. However, given that the connection among ROS nodes is based on simple TCP connections, it is possible to distribute the nodes on computers connected in a different way, as long as the lower-level communication is based on the TCP/IP (or UDP/IP) stack. This includes, of course, wireless networks.

However, as is well known, wireless communications are not as stable as their wired counterparts. In fact, the error rate is much higher, at least two order of magnitude in optimal conditions. This means that often  frames must be retransmitted several times at the MAC level (e.g. the default retransmission count is fixed at 7 for the 801.11 protocol in common Linux distributions) and this provokes a spreading of the communication end-to-end delays and jitter.

To illustrate the differences regarding the performance between distributed wired and wireless systems, we performed an experiment in which a distributed ROS system was simulated in both configuration. Two computers were directly connected first through Ethernet interfaces (100 Mbps) and then using two 802.11g wireless cards at 54 Mbps.
We generated a single flow with 64KB bytes messages and a period of 200 ms between them.

Figure \ref{fig:wired_vs_wireless} shows the distribution of the inter-arrival delay of the messages or, in other words, the period at the destination node. As expected, the figure shows a normal distribution around the original flow period of 200 ms.
However, as it can be seen, the wireless scenario presents a much higher degree of jitter showing a much wider normal distributions around the expected period. This, impacts the precision of the system and makes it more difficult to implement a high performance control loop, if needed.
Notice that the network was far from saturated in both cases,  as the required bandwidth of approximately 2.5 Mbps (plus TCP/IP overhead) was far below both the theoretical limit of the 802.11g protocol and, of course, the 100 Mbps of the Ethernet standard used.
This trivial example demonstrates what we should expect from wireless links:  higher jitter and thus less precise control of distributed systems. 

\subsection{Bandwidth limitations}
\label{sec:bandwidth}
Even as the bandwidths claimed by wireless standards are getting close to their wired counterparts, and sometimes even exceed them (the 802.11ac can theoretically reach a bandwidth of 3.67Gbps, for example), the performance in reality is often much worse. The claimed bandwidths are only reached when sender and receiver are close enough (e.g. the 802.11g standard guarantee a rate of 54 Mbps at a maximum distance of 37m indoor), the signal strength is high enough (generally this means they are in the line of sight of each other), and there are no interference, etc.
Furthermore, the real bandwidths are sometimes far from  theoretical ones, even in optimal conditions. The 802.11g standard, for example, has a maximum \textit{real} bandwidth (i.e. excluding MAC layer overhead) of about 36 Mbps \cite{jun2003theoretical}. 
Finally, and probably most importantly, the available bandwidth is reduced due to the impossibility of spatial reuse, inevitable in a large variety of situations. It can occur both due to the use of a wireless access point (AP) or in multi-hop ad-hoc communication.

In the first case, all the network nodes are in the range of communication of the AP and all the communication passes through it: schematically, the source send a frame to the AP that in turn forward it to the receiver. This means that only one node can transmit at a time otherwise the AP would lose the reception of the frames due to collision. Moreover, as explained, the propagation of the cited frames involves the exchange of two frames; this reduces the bandwidth to the half, like in a two-hops communication.

In the second case, something similar happens: if to extend the range of a hypothetical robot-team its members form a chain configuration (consider a search and rescue scenario in a long tunnel for example \cite{tardioli2015robot,Remley2008}), the frames exchanged by the robots in the two ends would need to hop through all the other relay nodes. This would reduce the available bandwidth by a factor $1/(n-1)$ in the worst case, being $n$ the number of nodes (computers) involved. Notice that even if in this situation some kind of spatial reuse could be possible, it strictly depends on the specific configuration. 

Bandwidth limitation is especially important in distributed ROS  systems: if different ROS nodes subscribe to some other ROS node's  topic, one network connection is established for each one of them. This could be especially dramatic if all connections take place over a wireless network, given that the bandwidth required is multiplied by the number of subscribers. Also, if the frequency a topic is published with is higher than that required, more bandwidth than the strictly necessary will be needed, jeopardising  global performance.
To be fair, some of these limitations can be overcome using the so-called \textit{topic\_tools} (\textit{relay} and \textit{throttle} nodes can do the trick) but they still don't offer a general and specific solution for wireless networked systems.
Another aspect that should be taken into account is the fact that, by default, ROS distributed systems rely on a centralized ROS core as explained above.
The handshakes necessary to connect different nodes ---also carried out establishing TCP connections--- add extra overhead that have a significant impact on the required bandwidth.

\begin{figure}[tpb]
\begin{center}
\includegraphics[page=1,width=1\columnwidth]{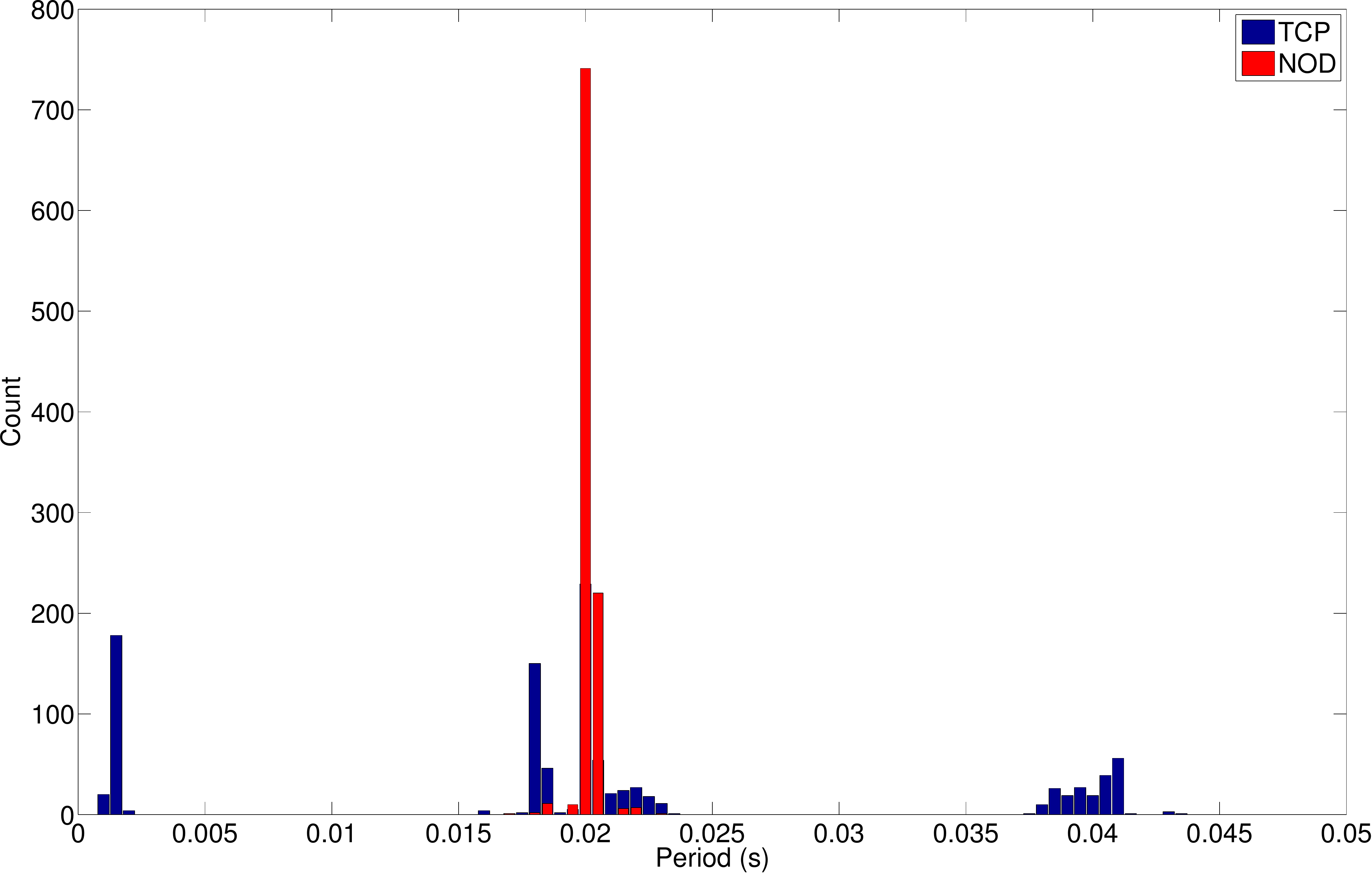}
\end{center}
\caption{\label{fig:nagle}\small Effect of the Nagle's algorithm.}
\end{figure}

\subsection{Delays due to Nagle's algorithm}
\label{sec:nagles}
The connection between ROS topics is performed using a TCP connection by default.
However, the Linux's implementation of the TCP protocol uses, again by default, the Nagle's algorithm to improve the efficiency of TCP/IP networks by reducing the number of packets sent over the network.
However, this can provoke delays in the transmission of the data that are buffered at the sender side before being transmitted. This means for example, that two ROS messages can be transmitted at the same time even if they are generated at different moments.
Unfortunately, this behavior takes place both in local and in over-the-network communication and can be extremely harmful in case of tight control loops.
Figure \ref{fig:nagle} shows a situation in which a single flow with 1KB messages and 20ms period is exchanged between two ROS nodes using two different connection options. The graph reports the period measured at the receiver side; as it can be seen TCP\_Nodelay (NOD in Figure~\ref{fig:nagle}) shows a narrow distribution around the source period while the TCPROS (TCP in Figure~\ref{fig:nagle}) shows three of them: around the real period, the double of the period and around 0 ms. 
As anticipated, this is due to the fact that, sometimes, two consecutive messages are buffered at the sender side and then sent together. 
At the receiver side they are, instead, published as soon as received through the socket.

\begin{figure}[tpb]
\begin{center}
\includegraphics[page=1,width=0.95\columnwidth]{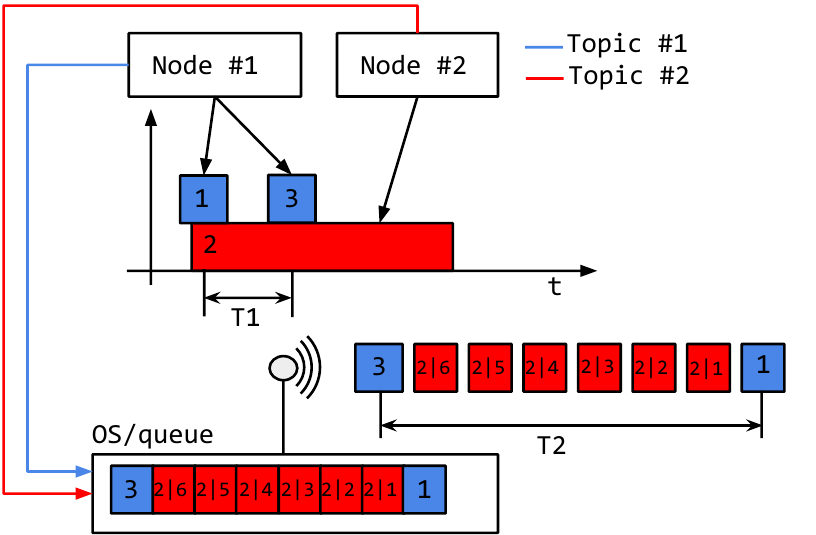}
\end{center}
\caption{\label{fig:priority_problem}\small Delay due to competing flows. The packet corresponding to message $\#3$ is delayed by packets corresponding to message $\#2$ ($\#2|1$ through $\#2|6$) in which it is split when entering the OS queue.}
\end{figure}

\subsection{Delays due to Competing flows}
\label{sec:competingflows}
Another aspects, the most important one in our view, is the performance of the communication in presence of competing flows.
This issue doesn't only appear in wireless networks but is more marked in this configuration due to the more limited bandwidth (especially in multi-hop networks, as described above) and the lossy nature of this medium.
When two or more flows have to share the same communication channel (either Ethernet or Wireless) the bandwidth must be shared as well, as is obvious. This should not be a problem if the flows don't saturate the network, but the transmission of one flow's messages can delay the transmission of the packets belonging to the other flow.

What happens in details is that when a node (or any application) sends any information (messages) through a socket, this is split in different MTU-sized packets (the MTU depends on the specific physical layer) and added to the FIFO operating system transmission queue. Since packets are then sent following an order in accordance with their position in the queue, any long messages 
introduce a noticeable delay in the transmission of the packets that are behind them in the queue. Consider figure \ref{fig:priority_problem}: the flow generated by \textit{Node \#1}, despite having a period $T1$ will see the packet corresponding to the messages published, sent with a period $T2>T1$ because a message coming from \textit{Node \#2} is pushed in the OS queue after message \textit{\#1} but before message \textit{\#3}. In other words message \textit{\#3} must wait for the transmission of packets \textit{\#2$|$1} through \textit{\#2$|$6} message \textit{\#2} is split into, being delayed noticeably.
Just to give an example, a \textit{64KB} message, that in a 802.11 \textit{6 Mbps} network with a $MTU=1500B$ is split in $\lceil65536/1500\rceil = 44$ radio packets, can introduce a delay of roughly $44\cdot(1500\cdot8)\cdot 10^{-6}/6 = 88ms$ (or $10ms$ in case of a \textit{54 Mbps} network).

\subsection{Blackouts due to Routing Issues}
Another aspect that should be treated, however is out of the scope of this paper, is the routing of the traffic. ROS itself does not offer any support for traffic routing that in case of wireless network must be performed by lower-layer protocols like  OLSR or BATMAN. However, these protocols have been proven very inefficient in case of node mobility provoking blackouts of several seconds inadmissible when it comes to tasks like robot control \cite{Haller2015}.

\subsection{Resilience}
Resilience to intermittent network disruptions is a desirable feature in any robotic system, particularly in field robots. Re-establishing the connectivity after disconnection is a core part of network resilience. Although ROS (TCP connection) can handle the network disruption effectively, it's important to know the how different protocols impact the efficiency of re-connection. In general, TCP connections are slower in getting reconnected, whereas UDP re-connections happen faster as they don't need handshaking process. For instance, this is the same reason why \textit{MOSH} \cite{mosh2012}, a remote terminal application, is preferable over the widely-used \textit{SSH}, for its support and robustness in dealing with intermittent connectivity.

\section{Solution Candidates}
\label{sec:solutions_analysed}

We need solutions to address the above problems and at the same time can also support multiple masters so that each computer (or robot) can perform its operation individually and hence be tolerant to communication disruptions or intermittent failures.  We found some candidate solutions in this direction.  

There are a few existing ROS tools that provide multiple ROS master support which offer a way of detecting other cores on the same or neighboring networks simplifying the development of multi-master systems. 
For instance, \textit{multimaster\_fkie} \cite{hernadez2015multi,Tiderko2016} helps to bridge multiple robots (computers) running ROS core in each of them but without actually improving the way the traffic is exchanged among the nodes. On the other hand, \textit{Nimbro}, \textit{Pound} and \textit{RT-WMP} propose a different, and in principle more efficient way of sending data from a source to a destination topic located in another computer.

Therefore, the candidate solutions considered in this paper are follows:
\begin{itemize}
    \item Standard ROS
    \begin{itemize}\small
    \item TCPROS (Referred as TCP in this paper, based on TCP)
    \item UDPROS (Referred as UDP in this paper, based on UDP)
    \item TCP\_NODELAY (also referred as NOD is this paper; it corresponds to TCPROS without the Nagle's algorithm)
    \end{itemize}
    
    \item \textit{Nimbro} network
    \item \textit{RT-WMP}
\end{itemize}

\accept{}{
Unlike standard ROS, the last two solutions, together with \textit{Pound} proposed here, rely on multiple \textit{ros-masters} (one per each node involved in the multi-robot team) and act similarly to a communication proxy, sending over the network only the actual ROS messages exchanged among the nodes encapsulating them in other lower-level protocols, as detailed in the next sections.}

There also exist some other ROS multi-master solutions (from individual researchers) such as \textit{multimaster\_experimental\footnote{\url{http://wiki.ros.org/multimaster_experimental}}}, \textit{wifi\_comm\footnote{\url{http://wiki.ros.org/wifi_comm}}} \cite{Kavan2012}, and \textit{multimaster-ros-pkg\footnote{\url{https://github.com/jonfink/multimaster-ros-pkg}}} however they are deprecated in the latest ROS versions, therefore not discussed here.

\subsection{ROS standard communication protocols}
TCPROS is the standard transport layer for ROS Messages and Services. It uses standard TCP/IP sockets for transporting message data. Inbound connections are received via a TCP Server Socket with a header containing message data type and routing information. 

The TCPROS method can be used with the \textit{NoDelay} option. If this option is set, Nagle's algorithm \cite{Nagle} is disabled. In this case the jitter should be reduced at the expense of a greater use of bandwidth.
Finally ROS allows the use of the \textit{unreliable} option that allows connectionless (acknowledgement-less) between different ROS nodes. In this case the protocol used is UDP but as the name suggests, there is no guarantees of message delivery.

\begin{figure}[tpb]
\begin{center}
\includegraphics[page=1,width=\columnwidth]{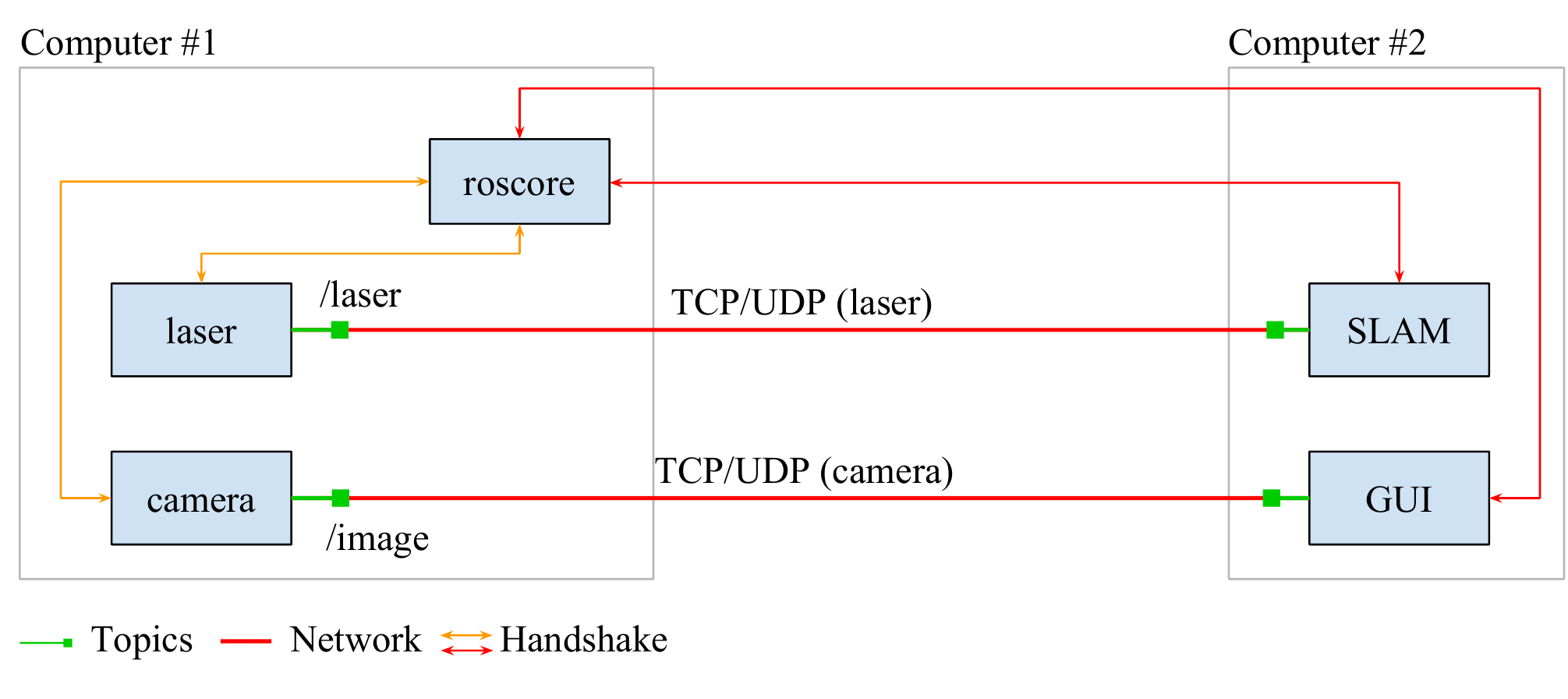}
\end{center}
\caption{\label{fig:schemes_ros}\small A hypothetical 2-nodes 2-topics system relying on standard ROS communication protocols. The LIDAR and camera ROS nodes publish data coming directly from the hardware while SLAM and GUI ROS nodes consume this information to create a map and to display the images respectively. One TCP or UDP connection is created for each topic after the handshaking managed by the only ROS core.}
\end{figure}

Figure \ref{fig:schemes_ros} shows the configuration that a hypothetical 2-nodes 2-flows system would have when this method of connection is used. A single \textit{roscore} manages the whole system and assists the connection between the two nodes as described above. One TCP or UDP independent connection\footnote{Although UDP is a connectionless protocol, we use the term UDP connection to refer to a UDP flow between two ROS nodes.} is established for each flow between source and destination

\subsection{\textit{Nimbro} Network}
\begin{figure}[t]
\begin{center}
\includegraphics[page=2,width=\columnwidth]{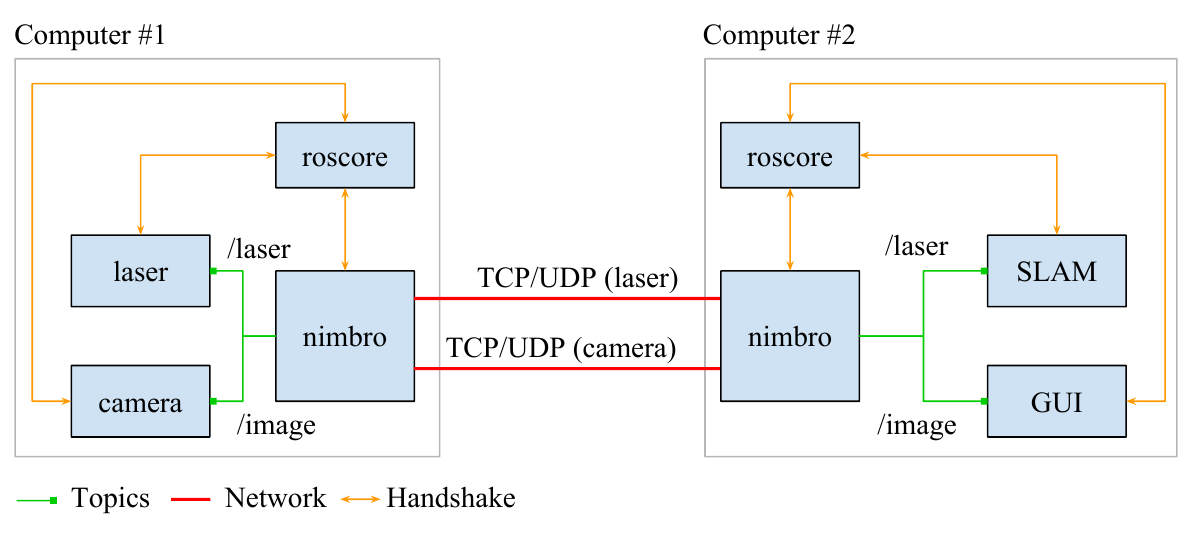}
\end{center}
\caption{\label{fig:schemes_nimbro}\small A hypothetical 2-nodes/2-topics system relying on \textit{Nimbro} nodes. One TCP or UDP connection managed by the \textit{Nimbro} nodes themselves is created for each topic. Local connection are managed by the ROS cores present in each computers.}
\end{figure}
This package, released by Bonn University is very flexible and offers several options. It has the possibility of \textit{transporting} topics (with the publication frequency being modulated at the transmission side) and services using TCP and UDP protocols using an optional transparent BZip2 compression to speed up communication and an experimental Forward Error Correction technique (based on OpenFEC) for UDP unreliable transport.
Additionally it provides nodes and filters for transmitting the ROS log, the TF tree and
H.264-compressed camera images. Finally, it includes an \textit{rqt} plugin for visualization and debugging of network issues.

Conceptually, the communication node subscribes to the the publishing topic, serializes the messages, and send it to the receiver side where it is deserialized and published.

\textit{Nimbro} allows easier configuration of the nodes and can be carried out by defining a simple configuration file which contains the settings such as the nodes to be transported, their maximum frequency, if compression should be used, the TCP/UDP port that the nodes should use, etc.

Again, Figure \ref{fig:schemes_nimbro} shows the configuration that a hypothetical 2-nodes/2-flows system would have when this method of connection is used. This time two independent ROS cores are used ---one for each computer--- and the connection between the topics on different computers is performed by the \textit{Nimbro} nodes themselves. However, like in the previous solution, an independent UDP/TCP connection is created for each topic to be transported.

In our set of experiments, we used the  $udp\_sender$, $udp\_receiver$, $tcp\_sender$, $tcp\_receiver$ nodes that, as their names indicate, use the UDP and TCP protocols respectively to transport the topics. 

\subsection{ROS \textit{RT-WMP}}
The \textit{RT-WMP} ROS nodes use a similar technique as \textit{Nimbro}, serializing the messages at the sender and recomposing them at the receiver side. However, they use the \textit{RT-WMP} protocol \cite{RT-WMP} as medium access and transport layer. This means, among other things, that the messages larger that the MTU, are split by the protocol itself into different network packets and sent independently.

\begin{figure}[t]
\begin{center}
\includegraphics[width=0.99\columnwidth]{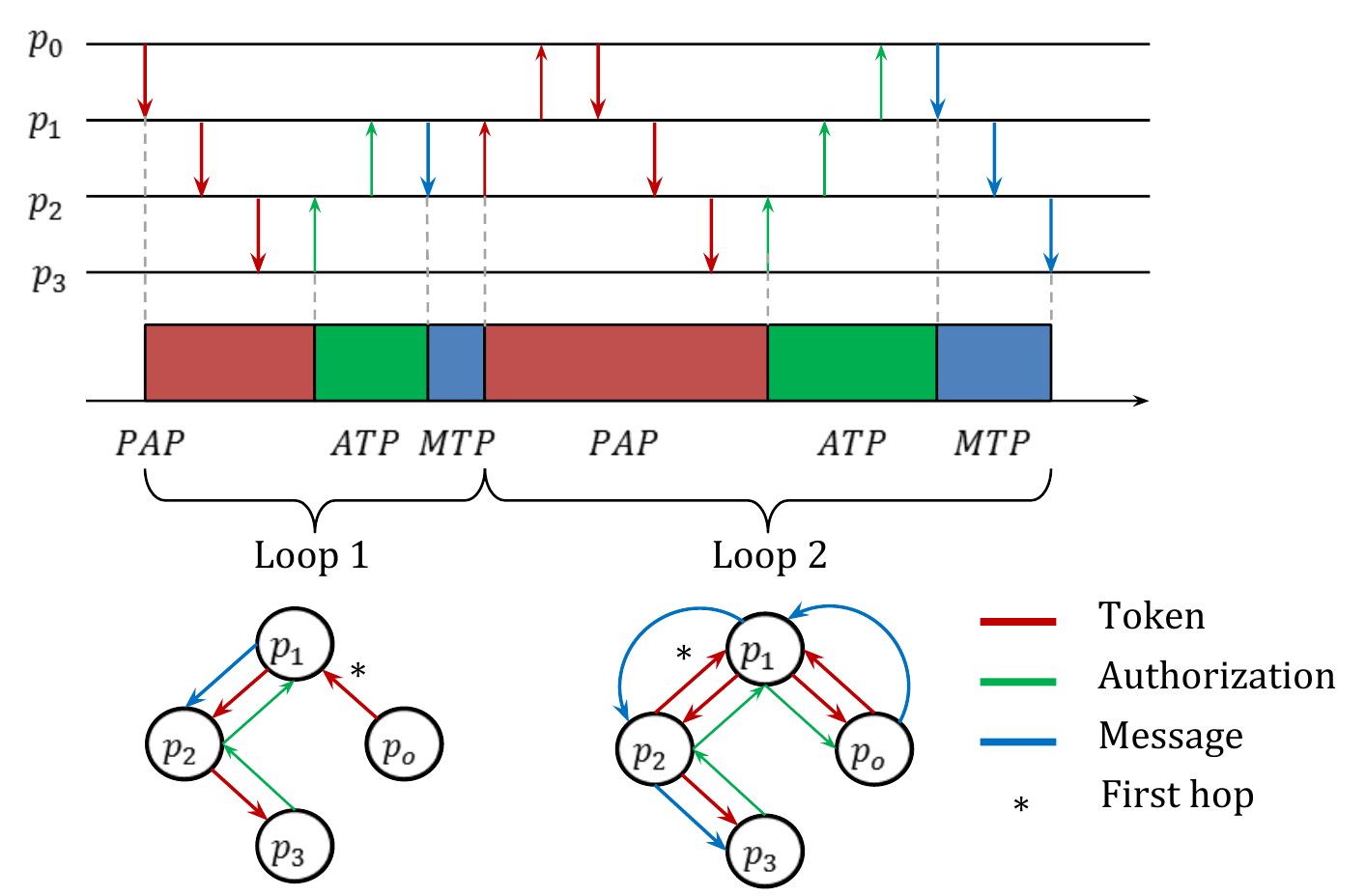}
\end{center}
\caption{\label{fig:rt-wmp-loops}\small Two different \textit{RT-WMP} loops.}
\end{figure}

The \textit{RT-WMP} is a token-based routing protocol that works with existing IEEE 802.11 networks. The protocol provides global static message priorities. Its target application is that of interconnecting a relatively small fixed-size group of mobile nodes, generally mobile robots (up to 32 units). It is based on a token-passing scheme and is designed to manage rapid topology changes through the exchange of a matrix containing the link quality among nodes. It works in three consecutive phases that repeat indefinitely (see Fig. \ref{fig:rt-wmp-loops}): the priority arbitration phase (PAP), the authorization transmission
phase (ATP), and the message transmission phase (MTP), together know as a loop.

During the PAP, nodes reach a consensus about which of them holds the Most Priority Message (MPM) at a particular moment. Next, 
in the ATP, an authorization is sent to the node that holds the MPM to authorize the transmission of the latter. 
Finally, in the MTP, this node sends the message to the destination. To reach a consensus over which node holds the highest priority, a token that holds information on the priority level of the MPM and on its owner among the set of nodes already reached, travels throughout the network during the PAP. 

The routing algorithm of the \textit{RT-WMP} is based on the link quality among nodes: a network connectivity graph having non-negative values on the edges describe the topology of the network. These values are computed as a function of the radio signal strength indicator (RSSI)  between pairs of nodes and are indicators of link quality between them. They are stored in the  so-called link quality matrix (LQM). Specifically, each column describes the links of a node with its neighbors. The nodes use this matrix to make decisions on the best path to route a message from a source to a destination (calculating a safe path using the Dijkstra algorithm). More details can be found in \cite{RT-WMP}.

\begin{figure}[tpb]
\begin{center}
\includegraphics[page=4,width=\columnwidth]{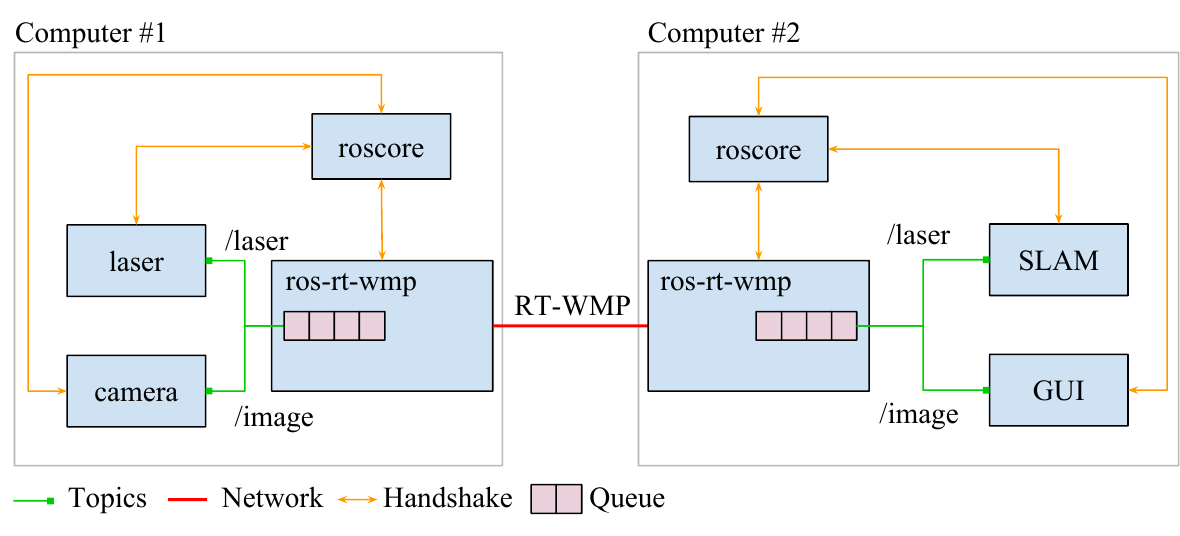}
\end{center}
\caption{\label{fig:schemes_rt_wmp}\small A hypothetical 2-nodes/2-topics system relying on \textit{RT-WMP} nodes. Messages are split in MTU-sized packets (if necessary) and pushed in a common queue ordered accordingly to the priority of the topic that originated the messages and sent over the channel in this order using the \textit{RT-WMP} protocol. Local connection are managed by the ROS cores present in each computer.}
\end{figure}

Figure \ref{fig:schemes_rt_wmp} shows that the configuration in case of  \textit{RT-WMP} is, in the practice, similar to the \textit{Nimbro} solution except that the packets/message are transported by the \textit{RT-WMP} protocol instead of TCP/IP or UDP/IP. 

\section{Proposed Solution: \textit{Pound}}
\label{sec:Pound}
In practical robot applications, each ROS flow (topic) has different latency and bandwidth requirements. Usually these two requirements conflict each other \cite{Parasuraman2013}. Recall from Section~\ref{sec:competingflows} that when two different type of flows compete each other for the same wireless channel, the latency or jitter requirements of both the flows may not be met. For instance, 
a laser topic could need to be exchanged with less delay compared to an image topic. Hence, priorities in various flows become vital in meeting the latency requirements of important flows and thereby maintain stability in the whole system.

We found that none of the candidate solutions (or even any existing ROS solutions) consider priorities for different flows except the \textit{RT-WMP} that, however, suffers from limitations due to the high overhead as will be shown. Therefore we propose a ROS node that implements priorities ---they can also be modified online---, can use compression, and can modulate the frequency of the source topics. We call it \textit{Pound} deriving from "Priorities Over UDP for reducing Network Delay". It offers less options to configure compared to \textit{Nimbro}, however we believe this make the \textit{Pound} easy to configure and use. Also it has the possibility of transporting the \textit{TF} topic excluding the information that are not relevant to the destination node. 

\begin{figure}[t]
\begin{center}
\includegraphics[page=2,width=0.95\columnwidth]{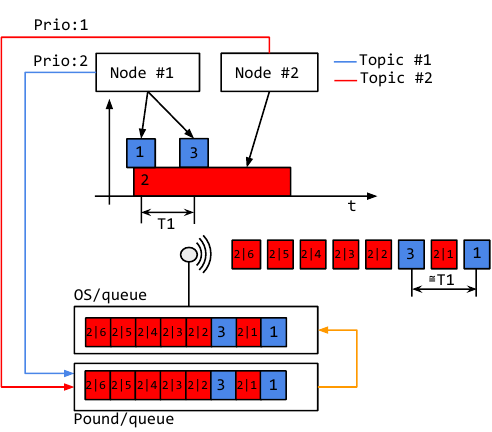}
\end{center}
\caption{\label{fig:priority_problem_pound}\small \textit{Pound} solution for competing flows with different level of importance. The messages enter the \textit{Pound} queue, are split in MTU-sized packets and the queue is sorted according to the priority of the topic that originated the messages. Then are pushed in the OS queue and later sent in the same order.}
\end{figure}

In order to implement the priority scheme proposed, the \textit{Pound} uses an intermediate transmission queue as shown in Figure \ref{fig:priority_problem_pound}. The communication is performed over a single UDP flow: the \textit{Pound} ROS node subscribes to the necessary topics, serializes the messages, split them in fragments that fits in a single radio packet and put them in a queue that is always maintained ordered according to the priority of the topic that generated each message. Then, it sends the packets following an order matching the priority of the source topics. On the other side, the messages are recomposed and published immediately. \accept{}{We used UDP as the communication means as it more likely supports real-time constraints than TCP.} 

Internally, the topics' callback functions fragment the messages and push them in the transmission queue indicating the priority associated with the topic itself; a transmission and a reception threads are in charge of popping the most priority message from the transmission queue and sending it  and listening for packets, respectively. At the reception side the messages are then reconstructed (when they are composed by multiple packets) and published. If some of the fragments is not received, the corresponding message is discarded. This means that there are not retransmissions at transport layer which simplifies the scheme and avoids congestion. Notice that this behavior is, in our opinion, the adequate in a real-time system given that late information is usually useless. 

As explained before, this solution is especially effective when dealing with topics that publish large messages given that small-sized higher priority messages do not need to wait for the transmission of larger messages having the opportunity to maintain their source period. 

\begin{figure}[tpb]
\begin{center}
\includegraphics[page=3,width=\columnwidth]{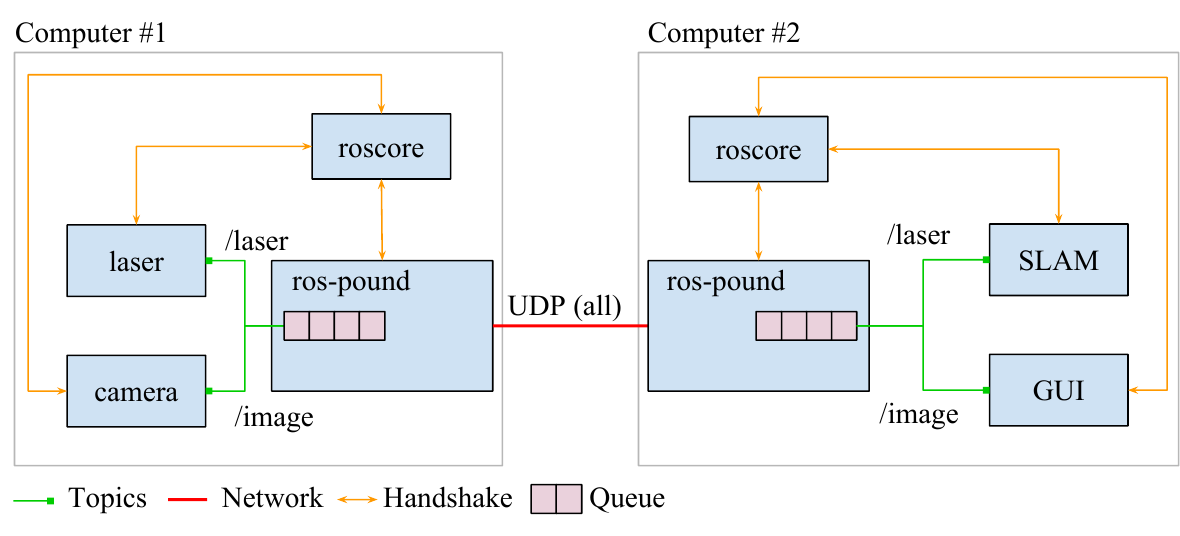}
\end{center}
\caption{\label{fig:schemes_pound}\small A hypothetical 2-nodes/2-topics system relying on \textit{Pound} nodes. A single UDP connection is created between source and destination. Messages are split in MTU-sized packets (if necessary) and pushed in a common queue ordered accordingly to the priority of the topic that originated the messages and sent over the channel in this order. Local connections are managed by the ROS cores present in each computer.}
\end{figure}

When transmitting, the \textit{Pound} node takes into account the time needed by the network to propagate the messages to avoid the congestion of the network.
For example, if the network rate is fixed at 6 Mbps, after sending a 1KB packet, the transmission thread will be put to sleep during $1024\cdot8\cdot 10^{-6}/6=1.3ms$ that corresponds roughly to the time needed to send the packet itself, before allowing it to pop the next packet from the \textit{Pound} transmission queue. Of course this value is not precise (also, the packet can be delayed before being transmitted if the channel is busy) but helps not to saturate the OS transmission queue and to reduce UDP discards. According to our experiments, this is particularly important when a large message is pushed into the queue because not doing so would imply a burst of consecutive calls to the socket \textit{send()} function that can cause the overflow of the OS transmission queue and, in turn, the discard of (some of the) packets making the reconstruction of the messages impossible at the receiver side.

The \textit{Pound} queue length can be limited to reduce network delay: when the queue is overflowing (which means that the bandwidth is not enough to allocate all the flows at the current rate), the node discard some of the messages starting from those coming from the less priority topics.

As shown in Figure \ref{fig:schemes_pound} and in the same way as \textit{Nimbro}, this solution uses multiples cores, one for each computer involved in the system.
However, this time a single UDP connection is used between source and destination and, as explained, the messages are sent according to the priority of the topic which published them.


\section{Evaluation}
\label{sec:experiments}

As explained in the introduction, the connection of ROS nodes over a wireless network can introduce serious delays and especially jitter if different flows are competing for the same communication channel. This is, in fact, the usual situation. Consider a search and rescue team of robots that are sent inside a tunnel to look for survivors as described in \cite{rizzo2013}. In this application, the network is composed by a set of robots that end up in a chain configuration and where the leader robot must send ---possibly over a multi-hop path--- sensor readings and camera images to the base station outside the tunnel while the joystick commands travel in the opposite direction. 

In the most basic setup, we have at least two competing flows (video feedback and LIDAR readings) in one direction and one flow (robot control commands) in the opposite direction competing for the available bandwidth.
All these flows, apart from using a different amount of bandwidth (the video feedback consumes more bandwidth than the other two), have different QoS requirements. If joystick commands arrive later than expected, for example, this can provoke a collision of the robot; 
in a similar way if the laser feedback suffer from a considerable delay, the human operator can send wrong commands with a similar outcome or with the undesirable consequence of system instability. 

Additionally, excessive jitter in the propagation of commands/feedback can make it difficult to close the control loop properly.
However, in this example system, not all the flows have same importance: it could be possible to reduce the frame rate of the camera images without losing usefulness, while this is not possible with the other two flows, at least while the robot is moving.

The test scenario we have chosen is inspired by the situation described above. We consider the two feedback flows ---\textit{laser} and \textit{image}--- to test the systems with one high-importance low-bandwidth flow and one high-bandwidth lower-importance flow sent simultaneously.

\subsection{Experiment setup}

The \textit{laser} flow has been configured to be composed in each run by 1000 messages of \textit{1KB} each with a 20ms period (50 Hz), while the \textit{image} flow consists of 500 messages of \textit{64KB} each with a 150ms period (6.67 Hz). The flows were generated using a dedicated ROS node called \textit{ros-profiler}\footnote{\url{https://github.com/dantard/unizar-profiling-ros-pkg/}} capable of publishing and subscribing to multiple flows with different message sizes and frequency.
With the methods that allowed it, we also tested different transport options for the two flows. Specifically, we tested a combination of UDP and NOD for the standard ROS protocol and UDP and TCP for the \textit{Nimbro} ---to which we refer to as \textit{Nimbro*}--- for \textit{laser} and \textit{image} flows respectively.

To summarize, the methods analysed in this paper and their identifiers are:
\vspace{5pt}
\begin{itemize}
    \item \textit{TCP}: Standard TCPROS protocol
    \item \textit{NOD}: Standard TCPROS protocol with No\_Delay option (Nagle's algorithm is off)
    \item \textit{UDP}: Standard UDPROS protocol
    \item \textit{UDP/NOD}: Standard ROS protocol in which the \textit{laser} flow was transported using ROSUDP protocol and the \textit{image} flow using the ROSTCP protocol
    \item \textit{Nimbro}: \textit{nimbro\_network} nodes in which both flows were transported using UDP protocol, \textit{udp\_sender} and \textit{udp\_receiver}
    \item \textit{Nimbro*}: \textit{nimbro\_network} nodes in which the \textit{laser} flow was transported using UDP protocol and the \textit{image} flow using TCP protocol, thus \textit{udp\_sender}, \textit{udp\_receiver} and \textit{tcp\_sender}, \textit{tcp\_receiver} respectively
    \item \textit{Pound}: \textit{pound} nodes in which all the flows were transported  using a single UDP connection. The \textit{laser} was assigned a higher priority than the \textit{image} flow
    \item \textit{RT-WMP}: \textit{ros-rt-wmp} nodes in which all the flows were transported using the \textit{RT-WMP} protocol. The \textit{laser} was assigned a higher priority than the \textit{image} flow
\end{itemize}
\vspace{5pt}

The different solutions have been tested using 802.11 a/b/g/n network Wi-Fi cards configured to work in a completely free channel in the \unit[5]{GHz} band. The data rate was fixed at \unit[6]{Mbps} to avoid automatic rate changes that could distort the measurements. This choice does not imply a loss of generality given that 1) similar behaviors can be noticed with higher network rates and higher required bandwidths, and 2) in large real-world environments robotics system often have to work with the lowest available rate. Similarly, we also decided not to use frame compression (available in \textit{Nimbro}, \textit{Pound} and \textit{RT-WMP}) given that 1) sometimes compression does not reduce significantly the size of the messages (e.g. JPG images), 2) compressed frames can be as large as 64KB or more, and 3) we want a fair comparison against standard ROS transport protocols.

The Wi-Fi cards were configured to work as ad-hoc peers and the measurements were performed in an 1-hop and 2-hops configuration (2- and 3-node chain networks respectively). Each network node is a computer running Ubuntu 14.04 OS and uses ROS Indigo version.
In the first case, the communication was peer-to-peer (Computer \#1 $\rightarrow$ Computer \#2) while in the second case we used a repeater configuring a standard IP forwarding and a fixed routing (Computer \#1 $\rightarrow$ Computer R $\rightarrow$ Computer \#2).
In case of the \textit{RT-WMP} the routing was also fixed providing the nodes with a fake LQM corresponding to the desired topology. 

\subsection{Measurements}
In our experiments, we analysed the jitter and delay of the message periods in both flows. 
The jitter was measured by calculating the inter-arrival delay at the receiver side or, in other words, the time span between the publication of a message and the subsequent one, which correspond with the period at the destination node.

To measure the delay we used the following technique: the messages are time-stamped at the sender side before being transmitted. When the destination node receives the message, it send back to the sender a frame containing such time-stamp through a dedicated Ethernet connection; in that moment the sender computes the elapsed time.
The values obtained in this way are not absolute given that they are affected by the delay introduced by the backward communication. However, this delay is practically constant and usually below 0.25 ms at the same time that it affects all the measurements in a similar way, thus allowing a fair comparison. Also, this avoids the problem of synchronizing computers clock which use to be a tedious and sometimes imprecise task due to clocks drift.

Additionally, we recorded two parameters that relates to the efficiency of the network connection. They are 1) \textit{Message Delivery Ratio (MDR)}: the ratio between messages successfully received and the total ROS messages sent over the network, and 2) utilized bandwidth (throughput) of the network. 


\subsection{Two nodes experiment}
In this experiment we used only two computers, one acting as the source and the other acting as the receiver.  

\subsubsection{Jitter}
Ideally, the recorded period values on the receiver should coincide with the message period. For instance, the \textit{laser} flow has a period of 20ms and hence the period at the receiver side should be the same. 

\begin{figure}[t]
\begin{center}
\includegraphics[page=1,width=0.95\columnwidth]{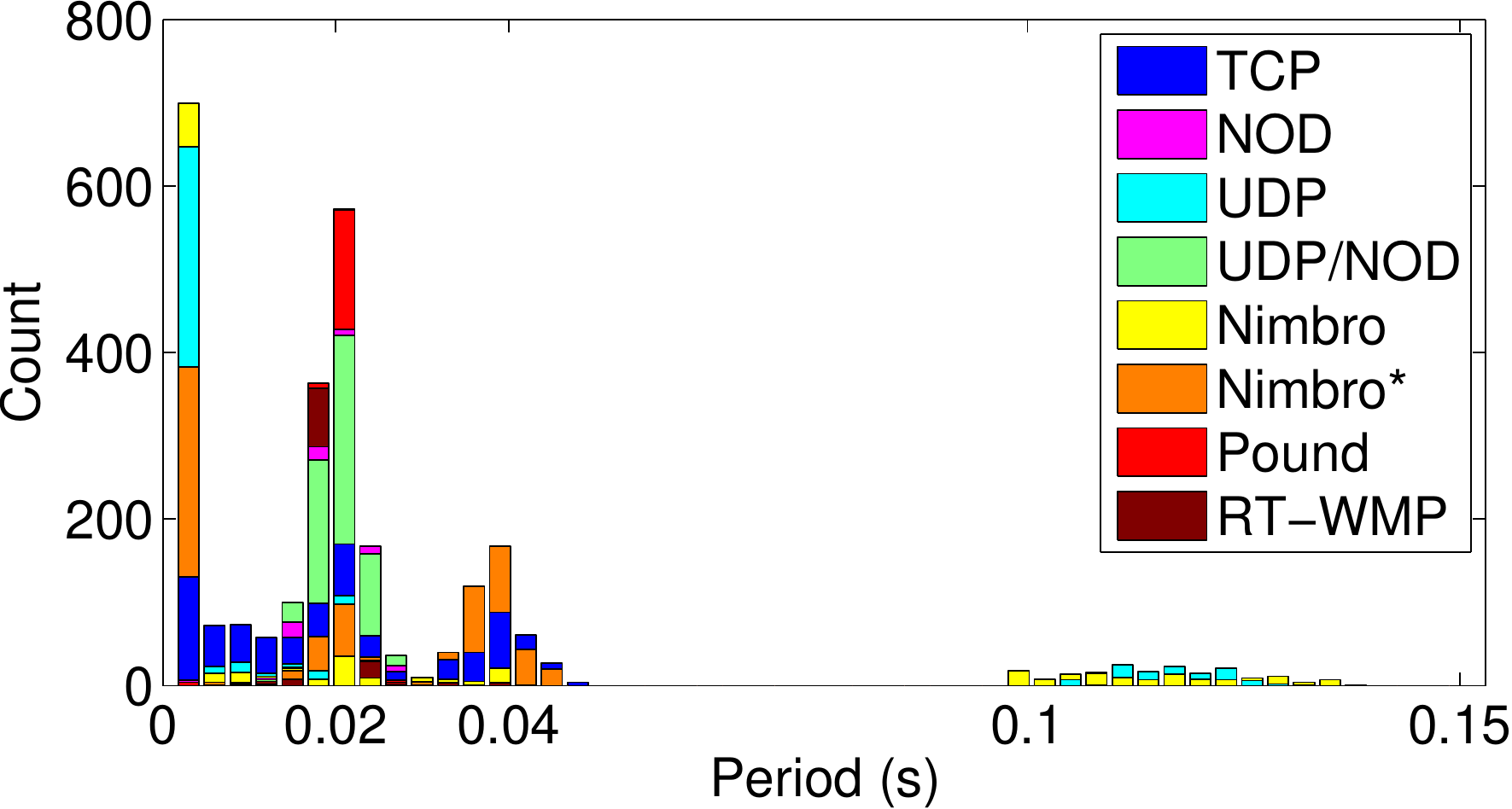}
\end{center}
\caption{\label{fig:jitter_laser}\small Jitter represented as variation of the period at the destination node for the \textit{laser} flow. In order to obtain a compact representation in this figure ---and in the others--- the bars are painted on top of each other in a way in which for each period value, the longest bar is painted first, then the second longest and so on.}
\end{figure}

Figure~\ref{fig:jitter_laser} shows the jitter period distribution for the different solutions analysed for the laser flow. As expected, the \textit{RT-WMP} and \textit{Pound} obtain the sharpest distributions around the real period, followed by the NOD and the UDP/NOD. The TCP connection shows different peaks as in the experiment shown in Section \ref{sec:nagles} around the real period, the double of the period (due to Nagle's algorithm) while \textit{Nimbro} presents a strange behavior with many messages published close to each other and other suffering from a jitter higher than 0.1 seconds. A similar thing happens with the UDP connection. The \textit{Nimbro*} behaves somehow better than the simple \textit{Nimbro} but also shows different peaks.
Notice that in this case the bandwidth required to allocate both flows is about 3.8 Mbps, below the theoretical limit of 802.11g when the rate is fixed at 6 Mbps.

\begin{figure}[t]
\begin{center}
\includegraphics[page=1,width=0.95\columnwidth]{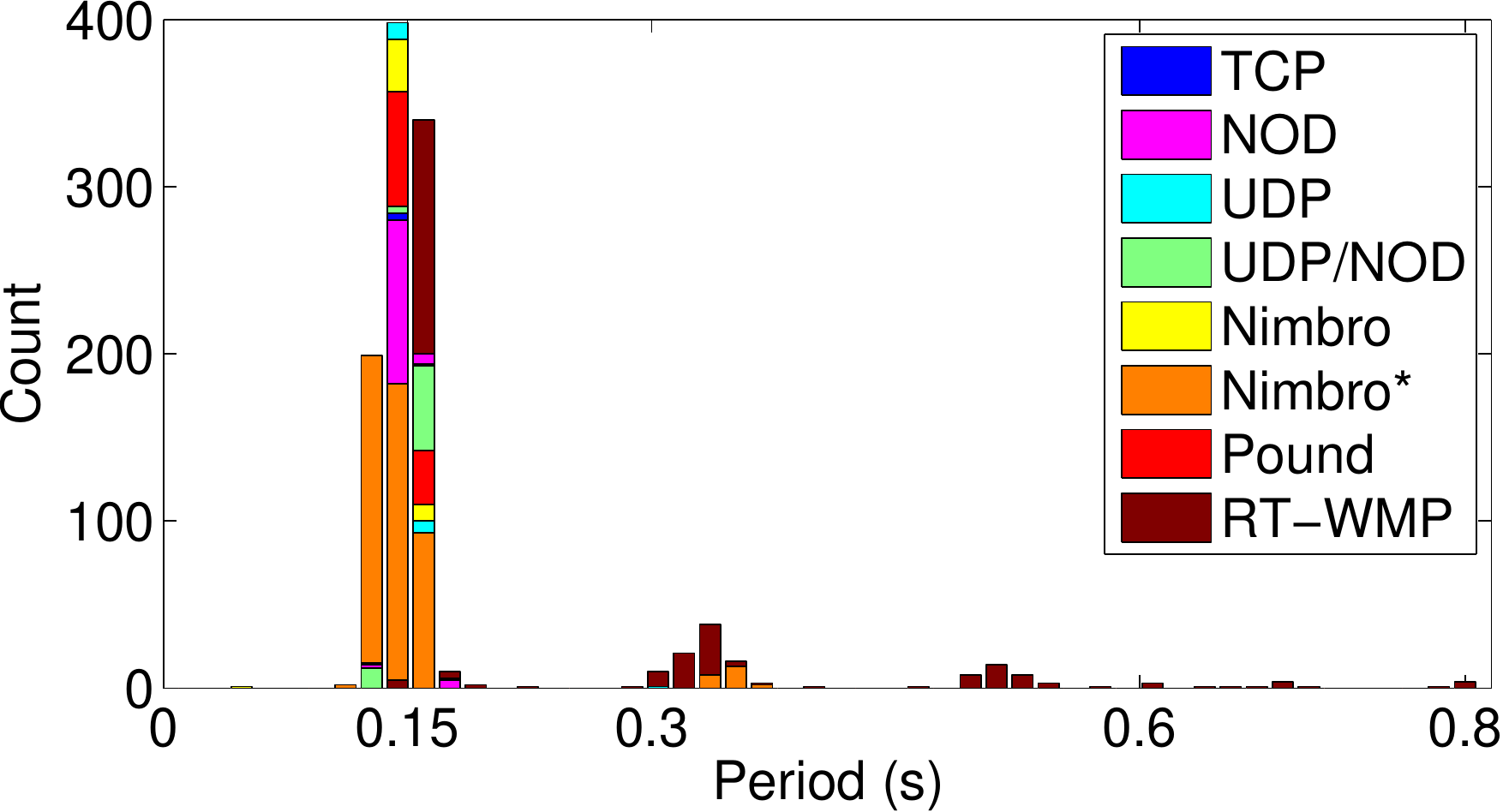}
\end{center}
\caption{\label{fig:jitter_image}\small Jitter represented as variation of the period at the destination node for the \textit{image} flow in a 2-nodes network.}
\end{figure}

Figure~\ref{fig:jitter_image} shows the jitter corresponding to the image flow. This time all the solutions show a distribution around the expected period. This is due to the fact that this flow's messages can only be delayed by just \textit{one} message of the other flow.
In this case, thus, \textit{Pound}, \textit{Nimbro} and UDP offer very similar and narrow distributions while NOD and TCP have similar behavior thanks to the fact that this time the messages of this flow are large enough not to need the Nagle's algorithm. On the other hand the \textit{RT-WMP} pays for its 3-phases delivering algorithm (which has much more overhead than the other methods) and the priority assigned to the message that, as expected, favour the \textit{laser} flow. The result is that image messages are delayed and sometimes discarded if the bandwidth offered in not enough to allocate both flows provoking a second distribution around the double of the period and over.


\begin{figure}
\begin{center}
\includegraphics[page=1,width=0.95\columnwidth]{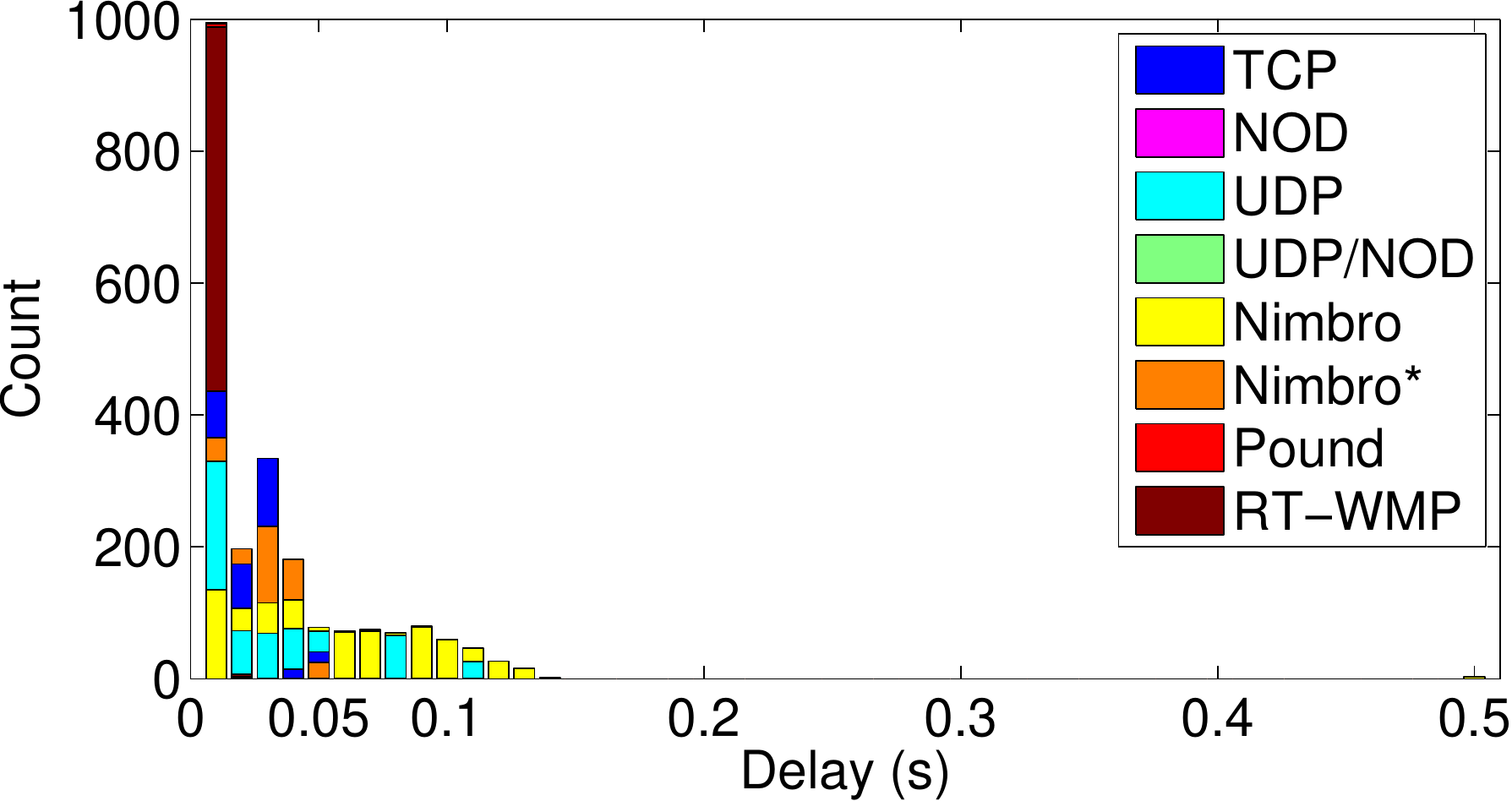}
\end{center}
\caption{\label{fig:delay_laser}\small Delay suffered by messages of the \textit{laser} flow in a 2-nodes network. Note: values for the \textit{Pound} appears close to and on top of \textit{RT-WMP}. }
\end{figure}
\begin{figure}
\begin{center}
\includegraphics[page=1,width=0.95\columnwidth]{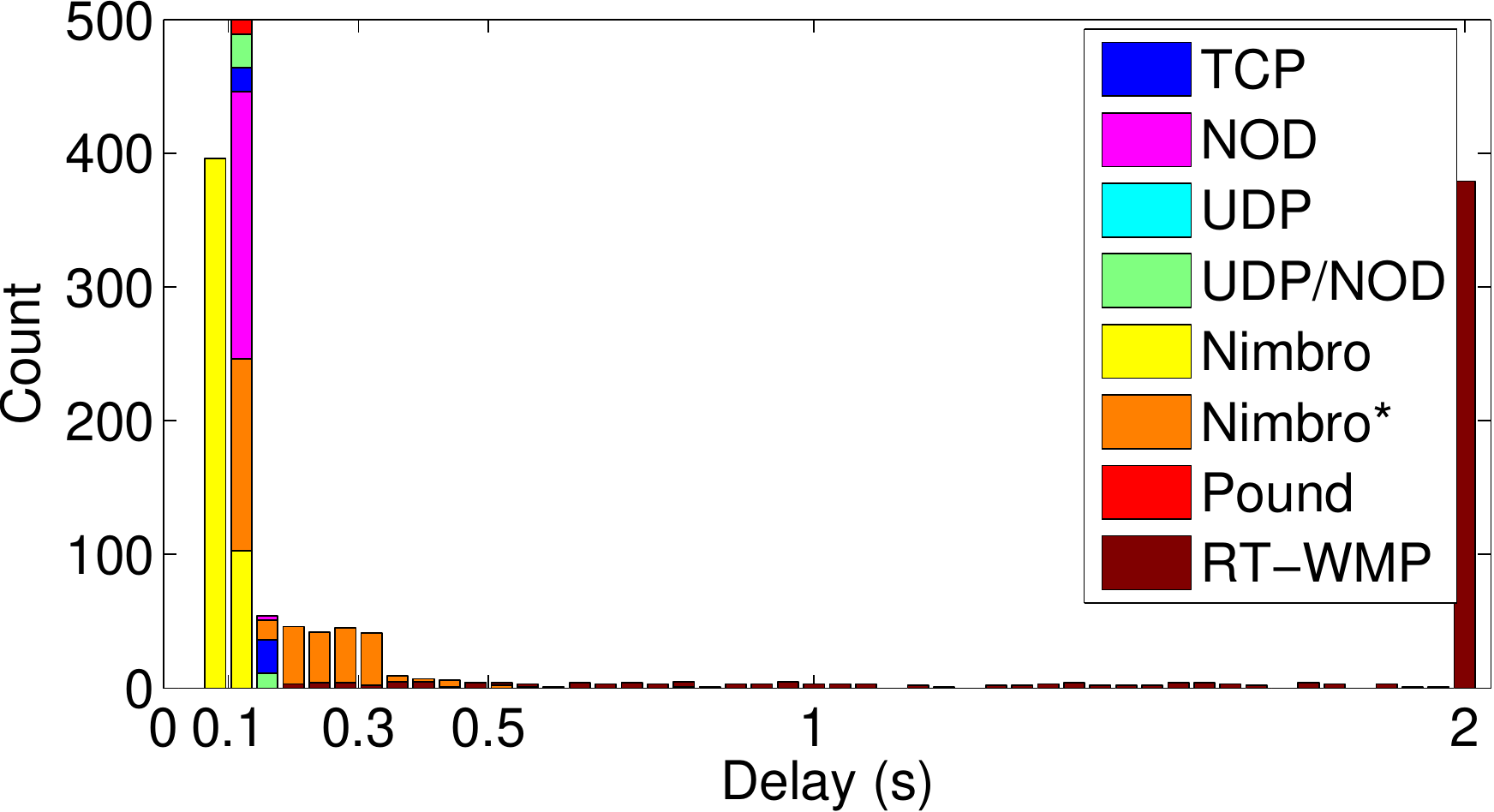}
\end{center}
\caption{\label{fig:delay_image}\small Delay suffered by messages of the \textit{image} flow in a 2-nodes network.}
\end{figure}
\subsubsection{Delay}

\begin{table}[]
\scriptsize
\centering
\caption{Laser topic. Results for 2-nodes network.}
\label{tab:laser_1h}
\resizebox{\columnwidth}{!}{%
\begin{tabular}{@{}l|rrrrrrr@{}}
& \multicolumn{1}{c}{$J_d$} & \multicolumn{1}{c}{$std(J_d)$} & \multicolumn{1}{c}{$D$} &\multicolumn{1}{c}{$std(D)$} &\multicolumn{1}{c}{$MDR$} &\multicolumn{1}{c}{$BW$} \\ 
& \multicolumn{1}{c}{(ms)} & \multicolumn{1}{c}{(ms)} & \multicolumn{1}{c}{(ms)} &\multicolumn{1}{c}{(ms)} &\multicolumn{1}{c}{(\%)} &\multicolumn{1}{c}{(Mbps)} \\ \hline
TCP     	 &      20.1 &	      13.3 &	      18.6 &	      11.6 &	  99.3	 &     0.398 \\ 
NOD     	 &      20.1 &	       4.2 &	       6.0 &	       6.5 &	  99.4	 &     0.398 \\ 
UDP     	 &      20.0 &	      38.2 &	      42.5 &	      34.8 &	  99.9	 &     0.400 \\ 
UDP/NOD 	 &      20.0 &	       3.2 &	       5.8 &	       4.0 &	  99.8	 &     0.399 \\ 
\textit{Nimbro}  	 &      20.0 &	      38.1 &	      55.8 &	      46.0 &	  99.8	 &     0.400 \\ 
\textit{Nimbro*} 	 &      20.1 &	      16.0 &	      22.2 &	      12.8 &	  99.3	 &     0.397 \\ 
\textit{Pound}   	 &      20.0 &	       2.0 &	       4.1 &	       4.0 &	  99.8	 &     0.400 \\ 
\textit{RT-WMP}  	 &      20.1 &	       2.8 &	       4.7 &	       5.0 &	  99.6	 &     0.399 \\ 
\end{tabular}
}
\end{table}

\begin{table}[]

\centering
\caption{Image topic. Results for 2-nodes network.}
\label{tab:image_1h}
\resizebox{\columnwidth}{!}{%
\begin{tabular}{@{}l|rrrrrrr@{}}
& \multicolumn{1}{c}{$J_d$} & \multicolumn{1}{c}{$std(J_d)$} & \multicolumn{1}{c}{$D$} &\multicolumn{1}{c}{$std(D)$} &\multicolumn{1}{c}{$MDR$} &\multicolumn{1}{c}{$BW$} \\ 
& \multicolumn{1}{c}{(ms)} & \multicolumn{1}{c}{(ms)} & \multicolumn{1}{c}{(ms)} &\multicolumn{1}{c}{(ms)} &\multicolumn{1}{c}{(\%)} &\multicolumn{1}{c}{(Mbps)} \\ \hline
TCP     	 &     150.1 &	       7.7 &	     132.3 &	       5.9 &	  99.9	 &     3.418 \\ 
NOD     	 &     150.1 &	       7.7 &	     132.7 &	       5.6 &	  99.9	 &     3.418 \\ 
UDP     	 &     150.4 &	       7.2 &	     106.3 &	       1.9 &	  99.7	 &     3.410 \\ 
UDP/NOD 	 &     150.1 &	       7.4 &	     129.1 &	       5.2 &	  99.9	 &     3.417 \\ 
\textit{Nimbro}  	 &     149.9 &	       5.3 &	     100.3 &	      31.5 &	 100.0	 &     3.421 \\ 
\textit{Nimbro}* 	 &     150.2 &	      44.7 &	     194.9 &	      86.2 &	  99.9	 &     3.416 \\ 
\textit{Pound}   	 &     150.2 &	       2.9 &	     112.0 &	       1.9 &	  99.9	 &     3.417 \\ 
\textit{RT-WMP}  	 &     233.6 &	     141.3 &	    4635.9 &	    2738.1 &	  64.2	 &     2.196 \\ 
\end{tabular}
}
\end{table}

Figure~\ref{fig:delay_laser} shows the corresponding delay distributions for the \textit{laser} flow. UDP and \textit{Nimbro} show a quite wide delay distribution (that is reflected also on the jitter at the receiver side, in fact) while NOD, \textit{Pound} and  \textit{RT-WMP} show a very constant delay below 10ms. Also UDP/NOD performs well, while \textit{Nimbro*} suffers from a delay that sometimes reaches 50ms. TCP, again due to Nagle's algoritm, show two different distributions with the second being around 25ms. Also, the graph reports a very small delay for \textit{RT-WMP} solution. This is due to the fact that in this configuration (i.e. with only two nodes on the network) each message transmission involves the exchange of 3 frames at most (usually two frames will be enough).

Figure~\ref{fig:delay_image} shows the corresponding delay distributions for the \textit{image} flow. In this case the most part of the flows show a delay between 100 ms and 130 ms except for \textit{Nimbro*} and \textit{RT-WMP}. The former shows a quite spread distribution with a mean of about 200 ms due to the fact that this flow, which uses the TCP transport, had to adapt to the bandwidth left free by the UDP flow. A similar situation happens with UDP/NOD but, surprisingly, this method shows a much better behavior.
The \textit{RT-WMP}, for its part, pays the limited bandwidth it can offer: the messages that are not discarded at the server side, are enqueued in its transmission queue and delivered with a delay that exceeds 4.5s.


Tables \ref{tab:laser_1h} and \ref{tab:image_1h} show corresponding numerical values of the mean period at the destination $J_d$, its standard deviation $std(J_d)$, the mean delay $D$, its standard deviation $std(D)$ the message delivery ratio $MDR$ and the Bandwidth consumed by the flow $BW$. All the mean jitter values are around 20ms but with great variability in terms of standard deviation. Something similar happen with the delay, with a mean that is in all the cases below 60ms but having a standard deviation that varies of up two orders of magnitude. On the other hand, all MDR are close to 100\% and all the methods consume, as expected, a similar amount of bandwidth except for the \textit{RT-WMP} that, being incapable of dealing with the required bandwidth, is forced to discard a noticeable amount of messages.

\subsection{Three nodes experiment}
This experiment is similar to the previous, except that we use an additional computer in between the source and receiver nodes to relay the information via wireless network. The relay node does the task of message forwarding between the source and receiver nodes. So, it is in-principle, a two-hop networks (i.e. a network with 3 nodes statically configured to form a chain). In this case the available bandwidth is virtually half compared to the previous experiment. For this reason, this time the two flows were configured with periods of 20ms (\textit{laser}) and 300ms (\textit{image}) respectively that correspond to a load slightly below the theoretical bandwidth offered by the wireless configuration used. Note the decrease in the message frequency of the \textit{image} flow.

The introduction of a relay node changes the situation significantly. On the one hand the probability of transmission error grows considerably over a 2-hop path and the same goes for the probability of a UDP (silent) discard. On the other hand TCP or even NOD protocols have not been designed to work over unreliable networks and the the scheme they use to retransmit lost packets can congest the network. 

\subsubsection{Jitter}
\begin{figure}[tpb]
\begin{center}
\includegraphics[page=1,width=0.95\columnwidth]{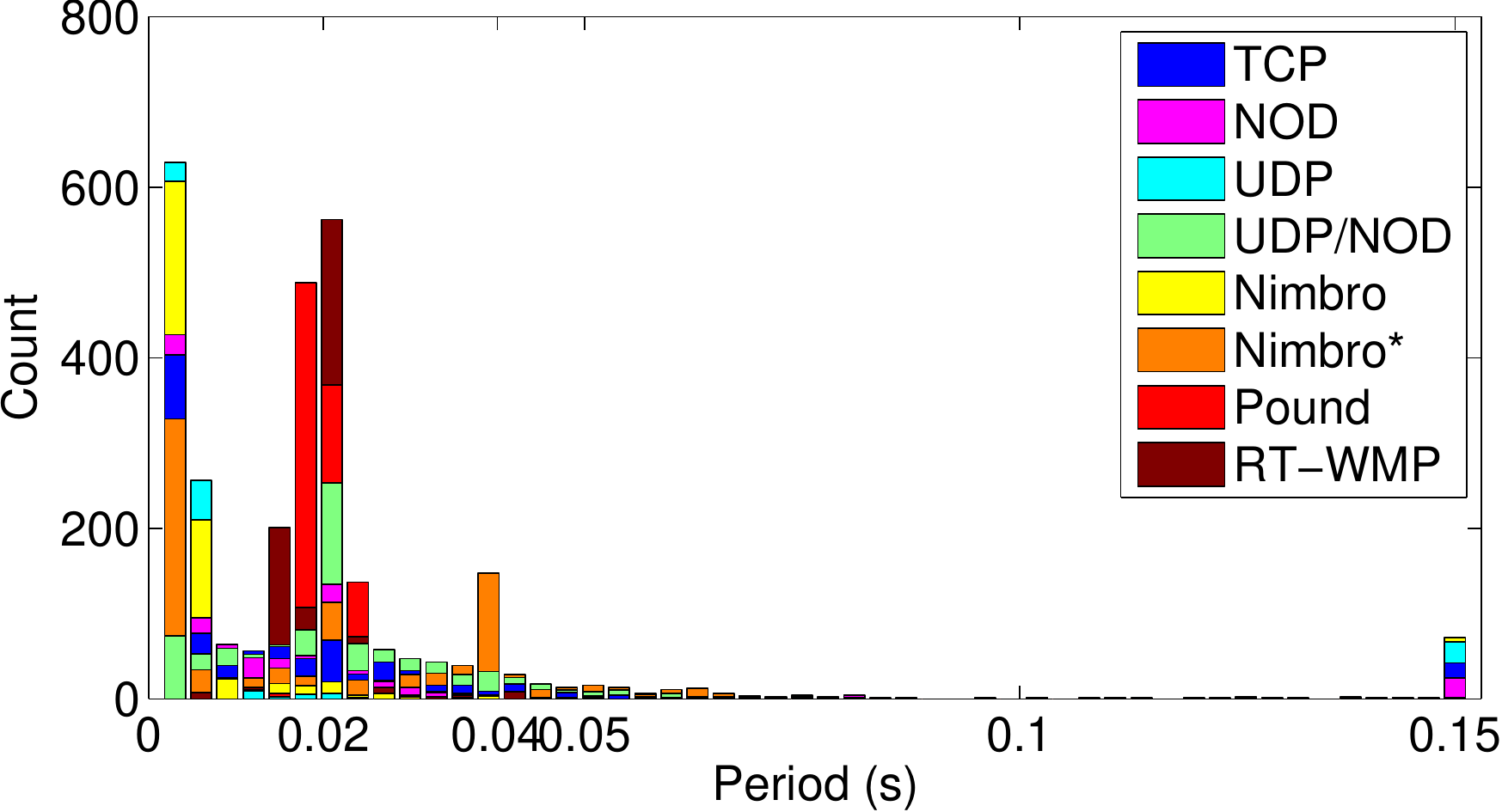}
\end{center}
\caption{\label{fig:jitter_laser_2h}\small Jitter represented as variation of the period at the destination node for the \textit{laser} flow in 3 nodes network}
\end{figure}

The above mentioned problems introduced by a relay node are reflected in Figure \ref{fig:jitter_laser_2h}: all the distributions are wider than before but additionally all the IP-based methods except \textit{Pound}, and somehow UDP/NOD combination (to a much lesser extent, though) are incapable of maintaining a sharp distribution around the period. This last combination (UDP/NOD) takes advantage of the fact that the TCP (NOD) protocol is designed to adapt itself to the available bandwidth left by the UDP flow (which is thus favored over the NOD flow). However what surprises the most is that the MDR for TCP and NOD are below 10\% and 40\% respectively despite being, supposedly, reliable protocols, showing a behavior in which the messages were delivered in bursts.
The \textit{RT-WMP}, meanwhile, shows two sharp distributions close to the expected period. This is due to the fact that in a 3-nodes networks the loops can vary more and the delivery of a message can take the transmission of 6 to 9 frames.

\begin{figure}[tpb]
\begin{center}
\includegraphics[page=1,width=0.95\columnwidth]{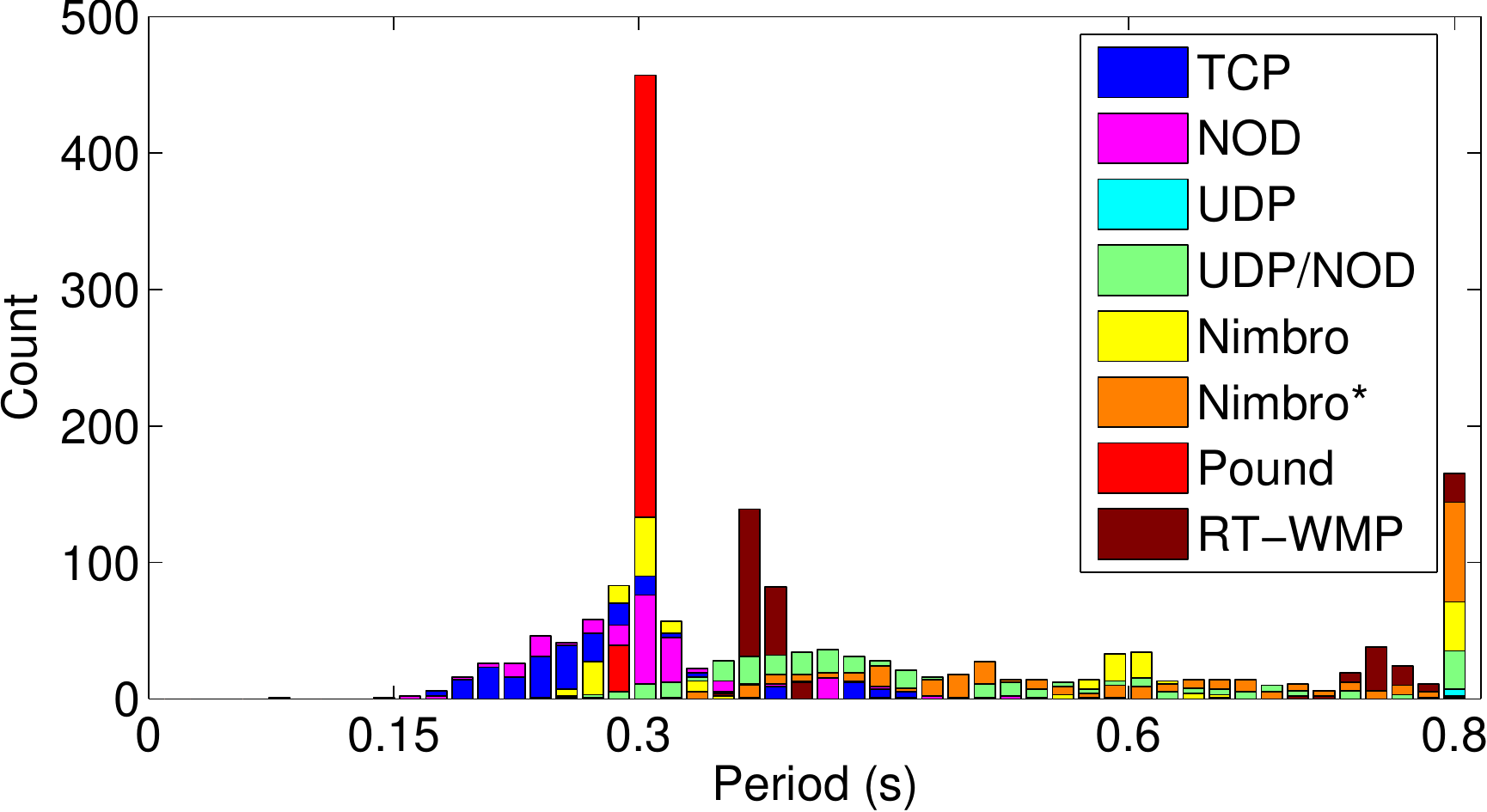}
\end{center}
\caption{\label{fig:jitter_image_2h}\small Jitter represented as variation of the period at the destination node for the \textit{image} flow in 3 nodes network. Several \textit{RT-WMP}, \textit{Nimbro} and \textit{Pound} messages showed a period above or equal to 0.6s mean.}
\end{figure}

Regarding the \textit{image} flow shown in Figure \ref{fig:jitter_image_2h} it is possible to see that only the \textit{Pound} shows a sharp distribution. \textit{Nimbro}, NOD and TCP also show a fairly good performance. Then there are UDP, UDP/NOD and \textit{Nimbro}* that in this experiment have been able to deliver just 4.91\%, 58.5\% and 38\% of the frames respectively. These two last methods pay for the fact that the \textit{image} flow uses TCP with a bandwidth limited by the UDP flow, that doesn't have any congestion detection method.
On the other hand, the \textit{RT-WMP} pays for its higher overhead and the fact of favouring the high priority flow being forced to discard about 73\% of the image frames. This is clearly reflected in the period on the destination node that presents a distribution at approximately 375ms and spread values beyond 0.8s (not visible in the figure).

\subsubsection{Delay}

In terms of delay for the \textit{laser} flow the \textit{Pound} is the method that performs better (having messages delayed by 6.8 ms on average) together with \textit{RT-WMP} with a delay below 10 ms. The measurement of the delay for TCP and NOD show a dual behavior with messages delivered within few milliseconds and other delayed several seconds. The UDP/NOD also perform well with a mean delay of 13ms. \text{Nimbro}* behave similarly but with higher delay and standard deviation. UDP and \text{Nimbro} instead show a quite delay spread with mean $\approx$120 ms.

\begin{figure}[tpb]
\begin{center}
\includegraphics[page=1,width=0.95\columnwidth]{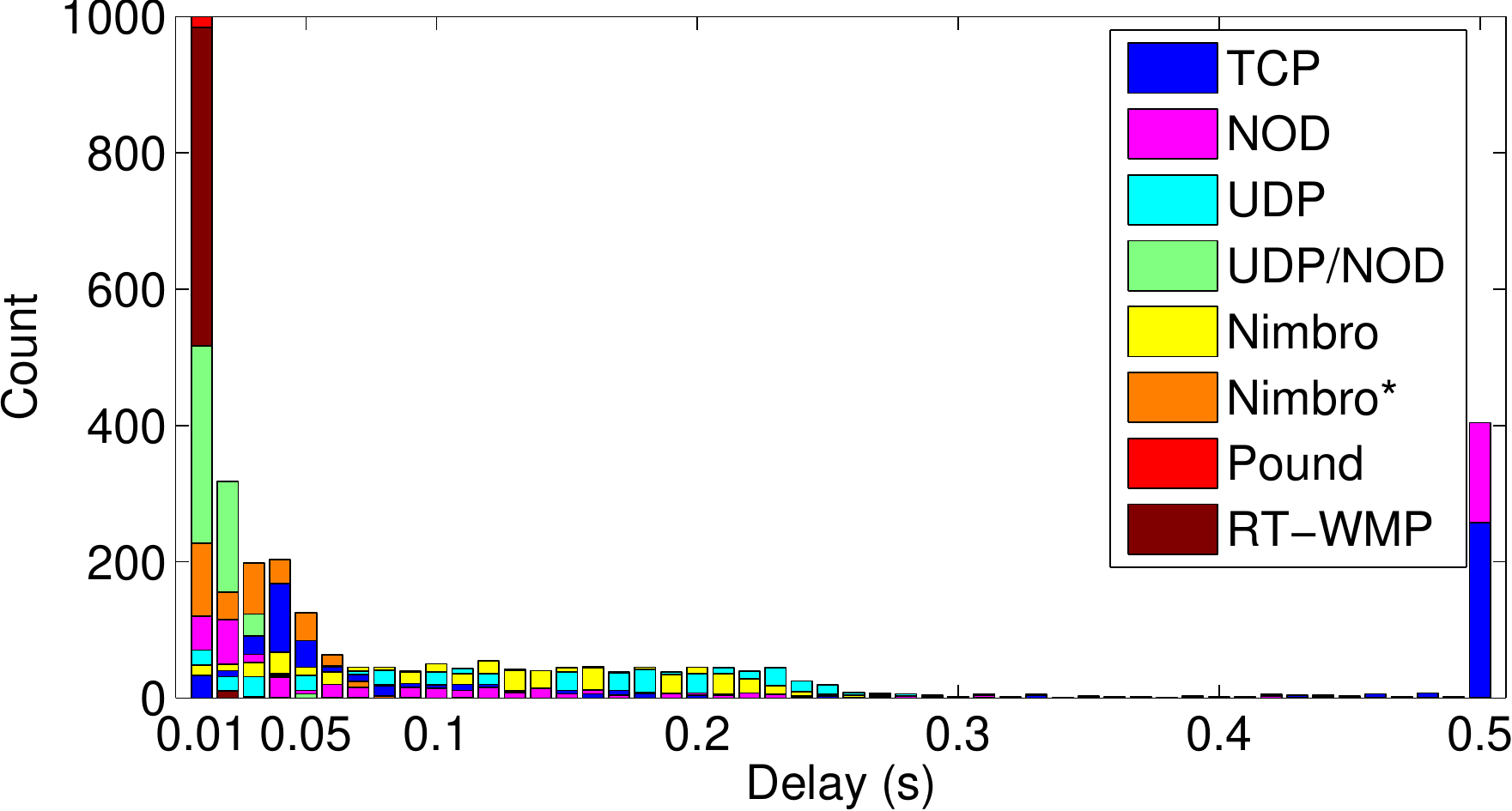}
\end{center}
\caption{\label{fig:delay_laser_2h}\small Delay suffered by messages of the \textit{laser} flow in 3-nodes network. Most TCP and NOD messages suffered a delay of $\approx$0.5s.}
\end{figure}

\begin{figure}[tpb]
\begin{center}
\includegraphics[page=1,width=0.95\columnwidth]{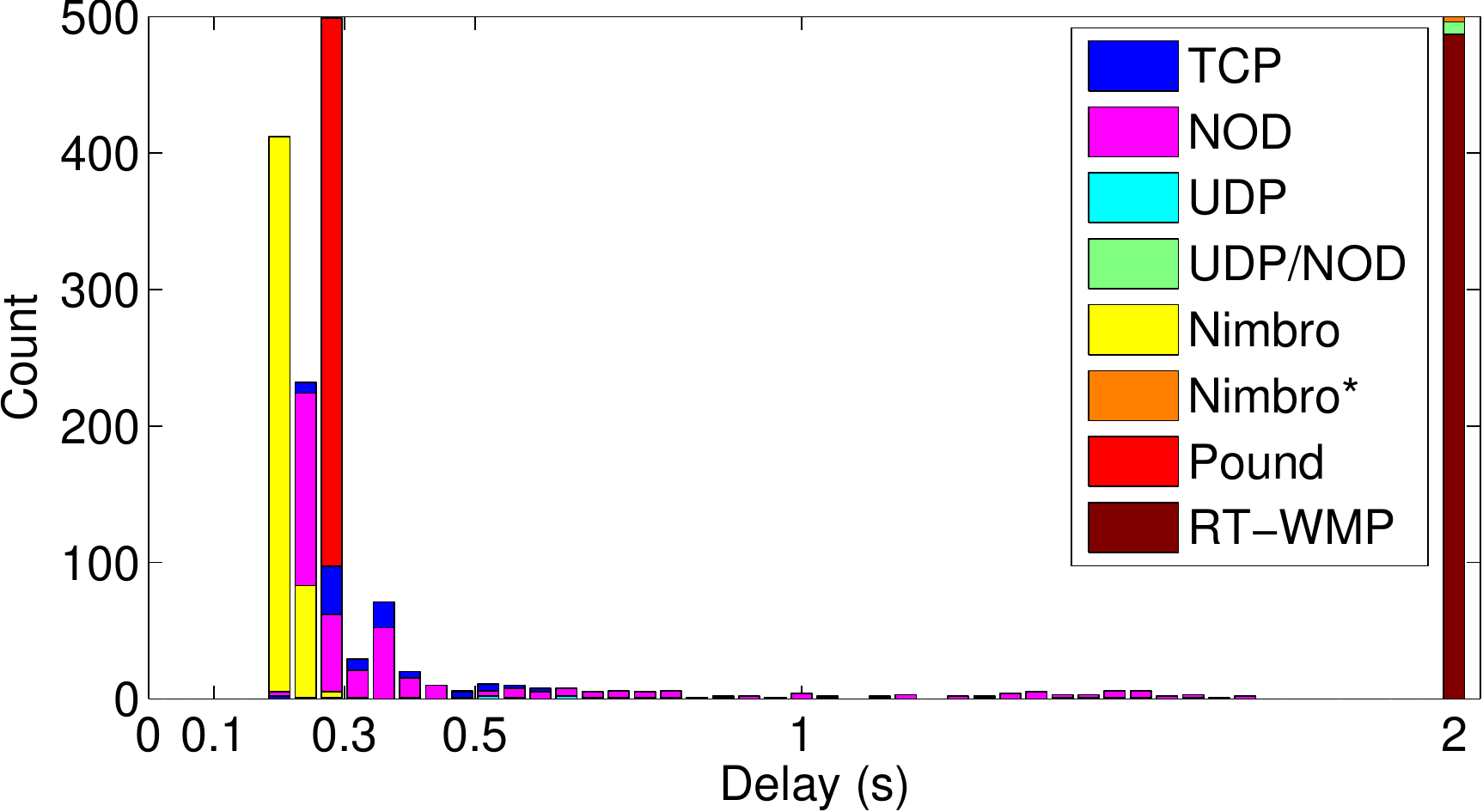}
\end{center}
\caption{\label{fig:delay_image_2h}\small Delay suffered by messages of the \textit{image} flow in 3-nodes network. All \textit{RT-WMP} messages suffered from a delay above 2s.}
\end{figure}

The situation is quite different in the \textit{image} flow where, this time, the \textit{Nimbro} shows the best behavior. \textit{Pound} shows a mean delay below 300ms in the same way as TCP but with a much lower deviation. Also NOD performs reasonably well while UDP/NOD and \textit{Nimbro}* show a significant delay: above 6 seconds and 27 seconds respectively.
Also \textit{RT-WMP} behaves extremely bad due to the fact that the high protocol overhead consumes a considerable part of the available bandwidth.

Tables \ref{tab:laser_2h} and \ref{tab:image_2h} show corresponding numerical values as in the previous experiment.

\begin{table}[t]
\centering
\caption{Laser topic. Results for 3-nodes network.}
\label{tab:image_2h}
\resizebox{\columnwidth}{!}{%
\begin{tabular}{@{}l|rrrrrrr@{}}
& \multicolumn{1}{c}{$J_d$} & \multicolumn{1}{c}{$std(J_d)$} & \multicolumn{1}{c}{$D$} &\multicolumn{1}{c}{$std(D)$} &\multicolumn{1}{c}{$MDR$} &\multicolumn{1}{c}{$BW$} \\ 
& \multicolumn{1}{c}{(ms)} & \multicolumn{1}{c}{(ms)} & \multicolumn{1}{c}{(ms)} &\multicolumn{1}{c}{(ms)} &\multicolumn{1}{c}{(\%)} &\multicolumn{1}{c}{(Mbps)} \\ \hline
TCP     	 &     257.1 &	    2870.8 &	    3127.6 &	   10243.4 &	   7.8	 &     0.031 \\ 
NOD     	 &      56.0 &	     787.3 &	    1201.7 &	    3869.0 &	  35.7	 &     0.143 \\ 
UDP     	 &      20.1 &	      61.7 &	     128.5 &	      74.6 &	  99.5	 &     0.398 \\ 
UDP/NOD 	 &      21.5 &	      11.7 &	      13.0 &	      10.4 &	  92.9	 &     0.372 \\ 
\textit{Nimbro}  	 &      21.6 &	      60.0 &	     113.1 &	      66.8 &	  92.4	 &     0.370 \\ 
\textit{Nimbro}* 	 &      21.4 &	     489.5 &	      41.9 &	     345.0 &	  93.6	 &     0.375 \\ 
\textit{Pound}   	 &      20.1 &	       1.7 &	       6.8 &	       1.4 &	  99.4	 &     0.398 \\ 
\textit{RT-WMP}  	 &      20.3 &	       4.2 &	       8.1 &	       4.5 &	  98.3	 &     0.394 \\ 
\end{tabular}
}
\end{table}

\begin{table}
\centering
\caption{Image topic. Results for 3-nodes network.}
\label{tab:laser_2h}
\resizebox{\columnwidth}{!}{%
\begin{tabular}{@{}l|rrrrrrr@{}}
& \multicolumn{1}{c}{$J_d$} & \multicolumn{1}{c}{$std(J_d)$} & \multicolumn{1}{c}{$D$} &\multicolumn{1}{c}{$std(D)$} &\multicolumn{1}{c}{$MDR$} &\multicolumn{1}{c}{$BW$} \\ 
& \multicolumn{1}{c}{(ms)} & \multicolumn{1}{c}{(ms)} & \multicolumn{1}{c}{(ms)} &\multicolumn{1}{c}{(ms)} &\multicolumn{1}{c}{(\%)} &\multicolumn{1}{c}{(Mbps)} \\ \hline
TCP     	 &     301.0 &	      68.9 &	     303.6 &	      95.8 &	  99.7	 &     1.704 \\ 
NOD     	 &     300.0 &	      88.2 &	     431.6 &	     347.6 &	 100.0	 &     1.710 \\ 
UDP     	 &    6908.4 &	    9351.2 &	     506.7 &	     160.1 &	   4.3	 &     0.085 \\ 
UDP/NOD 	 &     512.6 &	     207.0 &	   27218.3 &	    7530.2 &	  58.5	 &     1.001 \\ 
\textit{Nimbro}  	 &     486.6 &	     340.5 &	     209.1 &	      13.7 &	  61.7	 &     1.054 \\ 
\textit{Nimbro}* 	 &     788.8 &	     865.1 &	    6618.9 &	     869.1 &	  38.0	 &     0.650 \\ 
\textit{Pound}   	 &     300.7 &	      13.6 &	     294.2 &	       2.8 &	  99.8	 &     1.706 \\ 
\textit{RT-WMP}  	 &    1108.2 &	    1153.9 &	   14479.1 &	    4484.5 &	  27.1	 &     0.463 \\ 
\end{tabular}
}
\end{table}

\subsection{Bandwidth}
\begin{figure}[b]
\begin{center}
\includegraphics[page=1,width=0.85\columnwidth]{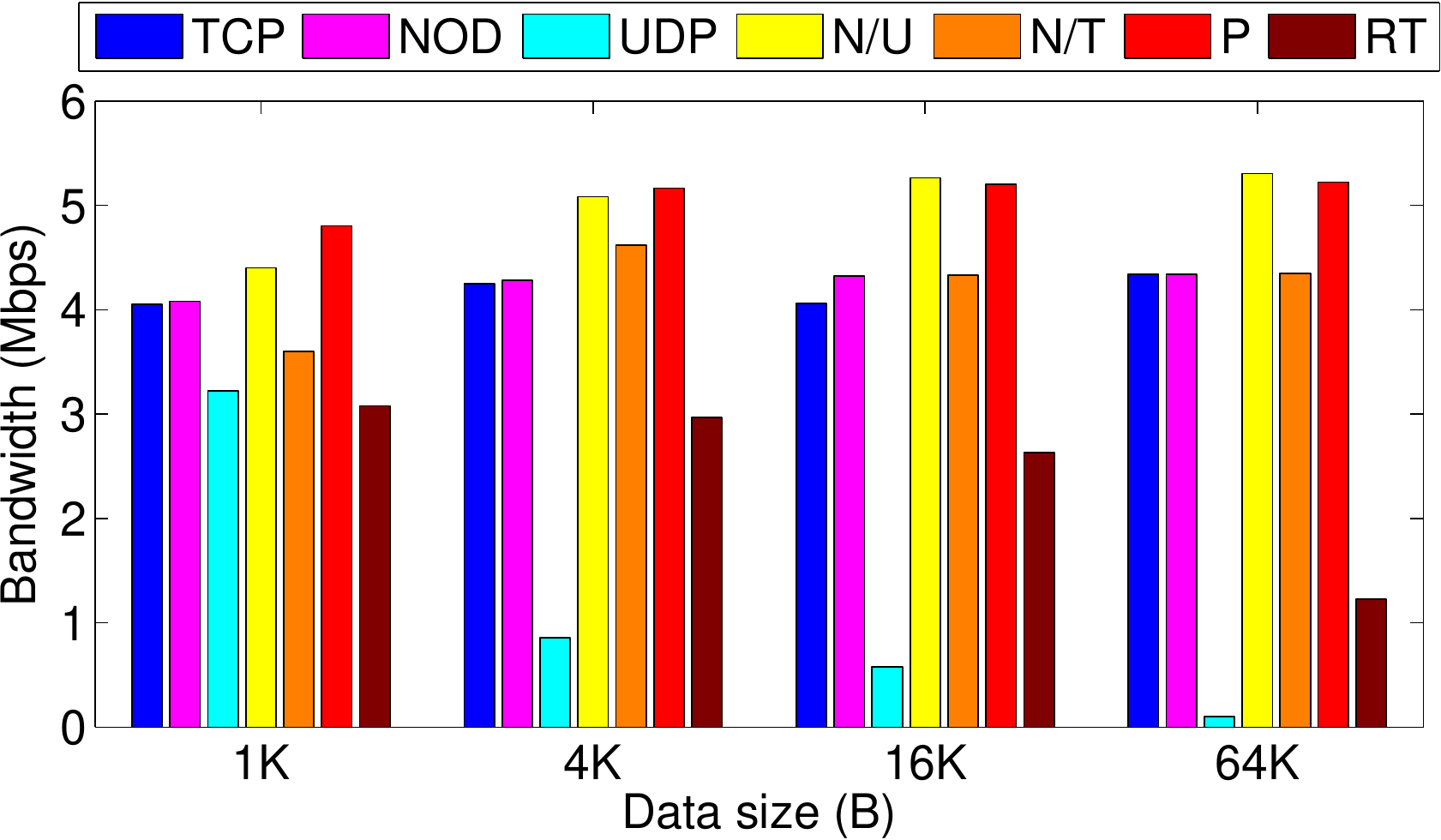}
\end{center}
\vspace{-2mm}
\caption{\label{fig:bandwidth}\small 
Bandwidth offered by the different methods. \textit{N/U} stands for \textit{Nimbro/UDP}, \textit{N/T} for \textit{Nimbro/TCP}, \textit{P} for \textit{Pound} and \textit{RT} for \textit{RT-WMP}.
}
\end{figure}

Figure \ref{fig:bandwidth} shows the user-bandwidth offered by the methods evaluated in a 2-nodes network configured in the same way as the previous experiments. This has been computed considering saturated traffic at the sender side (of $\approx$6.5 Mbps) and for different sizes of the messages. This time a single flow was transported. The results show that the \textit{Nimbro/UDP} and \textit{Pound} methods provide the best performance with similar and constant bandwidths, above 5 Mbps for all the different message sizes. TCP, NOD and \textit{Nimbro/TCP} also offer good (and approximately constant) results close to 4 Mbps; the difference is due to the extra overhead that the TCP protocol has with respect to UDP. 
On the other hand UDP shows a performance that clearly decreases with the size of the message due to the growing amount of discarded packets that cause an almost vanishing bandwidth for messages of 64KB.
Finally, \textit{RT-WMP} provide a decreasing bandwidth: large messages need the exchange of a higher number of packets and given that for each message the protocol executes a 3-phase consensus algorithm, the overhead grows considerably with the size of the message to be sent limiting the available user bandwidth.
It is worth remarking, however, that a higher raw bandwidth does not necessarily mean a better result if it is not provided guaranteeing a limited amount of jitter and delay, especially in a system with tight control loops.

\begin{figure}[t]
\begin{center}
\includegraphics[page=1,width=0.9\columnwidth]{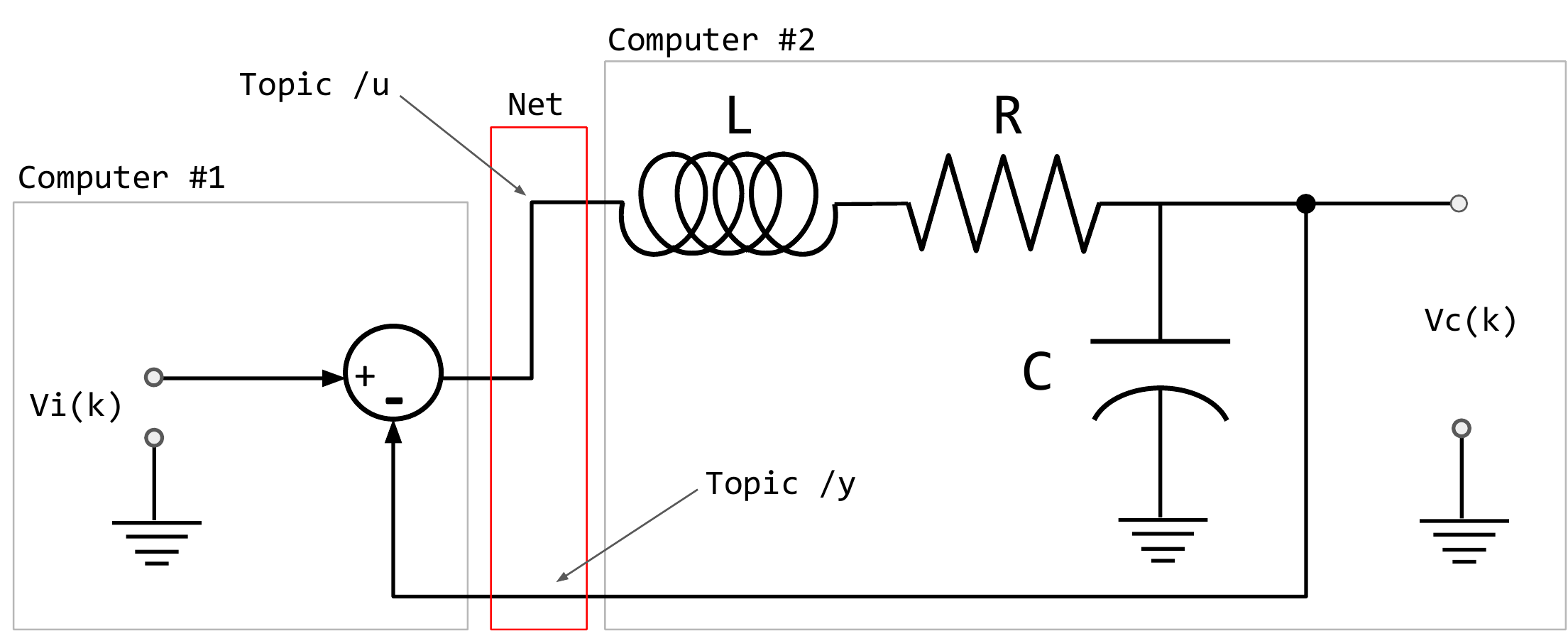}
\end{center}
\vspace{-2mm}
\caption{\label{fig:rlc}\small RLC control system used in the experiment.}
\vspace{-2mm}

\end{figure}

\subsection{Resilience}
As anticipated earlier, another important aspect to take into account in robotics systems is the resilience against unexpected situations that can take place, for example, in search and rescue scenarios. To check one of the many aspects involved in these situations, we performed a simple experiment in which the intermittent operation of a network node is simulated.
We set up, in the usual 2-nodes and 3-nodes chain configuration, a scenario in which 2 flows, both with 20ms period and 1KB payload, were transmitted between the head and the tail of the chain. Then we simulated the malfunctioning of the sender node (in the 2-nodes experiment) or of the relay node (in the 3-nodes experiment) bringing down the interface for exactly 5 seconds and then bringing it up again.
The goal was to measure the time the different methods take to reestablish the flows of data after this issue.
Notice that this includes the time needed by the wireless cards to rejoin the ad-hoc network. The experiment is thus not absolute but must be interpreted as a basic comparison among the different methods.

The results are shown in table \ref{tab:downup}. From the data it can be concluded that, as expected, the methods based on UDP are about half a second faster that TCP ---it has to renegotiate the connection--- and that there is a difference of around 300 ms between the 2-nodes and 3-nodes systems.
On the other hand the \textit{RT-WMP} that doesn't use the IP protocol but RAW 802.11 frames, shows a much faster behavior performing the switch in about 0.5 seconds.

\begin{table}[h]
\centering
\caption{Reconnection time after node failure (s)}
\label{tab:downup}
\begin{tabular}{@{}l|cc|cc@{}}
& \multicolumn{2}{c|}{2-nodes} & \multicolumn{2}{c}{3-nodes}  \\ 
& \multicolumn{1}{c}{Flow 1} & \multicolumn{1}{c|}{Flow 2} & \multicolumn{1}{c}{Flow 1} & \multicolumn{1}{c}{Flow 2} \\ \hline
TCP     	 &     1.73 &    1.93 &    2.19 &    2.28\\
NOD     	 &     1.96 &    1.96 &    2.46 &    2.32\\
UDP     	 &     1.29 &    1.28 &    1.47 &    1.46\\
UDP/NOD 	 &     1.28 &    1.92 &    1.46 &    2.32\\
\textit{Nimbro}  	 &     1.31 &    1.30 &    1.46 &    1.46\\
\textit{Nimbro}* 	 &     1.32 &    1.68 &    1.64 &    2.99\\
\textit{Pound}   	 &     1.31 &    1.30 &    1.53 &    1.52\\
\textit{RT-WMP}  	 &     0.42 &    0.54 &    0.50 &    0.49\\
\end{tabular}
\end{table}

\subsection{Control loops}
\label{sec:control-loops}

The last test investigated the capabilities of the methods to deal with a real control loop in presence of another \textit{perturbing} flow.
We set up a system like the one shown in Figure \ref{fig:rlc} implementing the (corresponding discrete-time) system itself in a computer and the (corresponding discrete-time) controller (in this case just a P controller with $K=1$) in another computer. The two parts were connected through two ROS topics (namely $u$ for the control signal and $y$ for the feedback) transported from a computer to another with some of the methods being analysed in this paper.
The values were fixed as $C=0.1, L=0.1$ and $R=1$ to obtain an under-damped system with a transient response of about 2 seconds and an overshoot of about 40\%. The period was fixed in $T=20ms$ and $Vi(k)=1V$. The $u$ and $y$ flows (about 4Kbps each) were perturbed by another flow from \textit{Computer \#2} to \textit{Computer \#1} having period of $200ms$ and message-size of $64KB$ (\text{2.5 Mbps}, approx). In case of \textit{Pound} and \textit{RT-WMP}, $u$ and $y$ were given higher priorities than the perturbing flow.

\begin{figure}[t]
\begin{center}
\includegraphics[page=1,width=0.85\columnwidth]{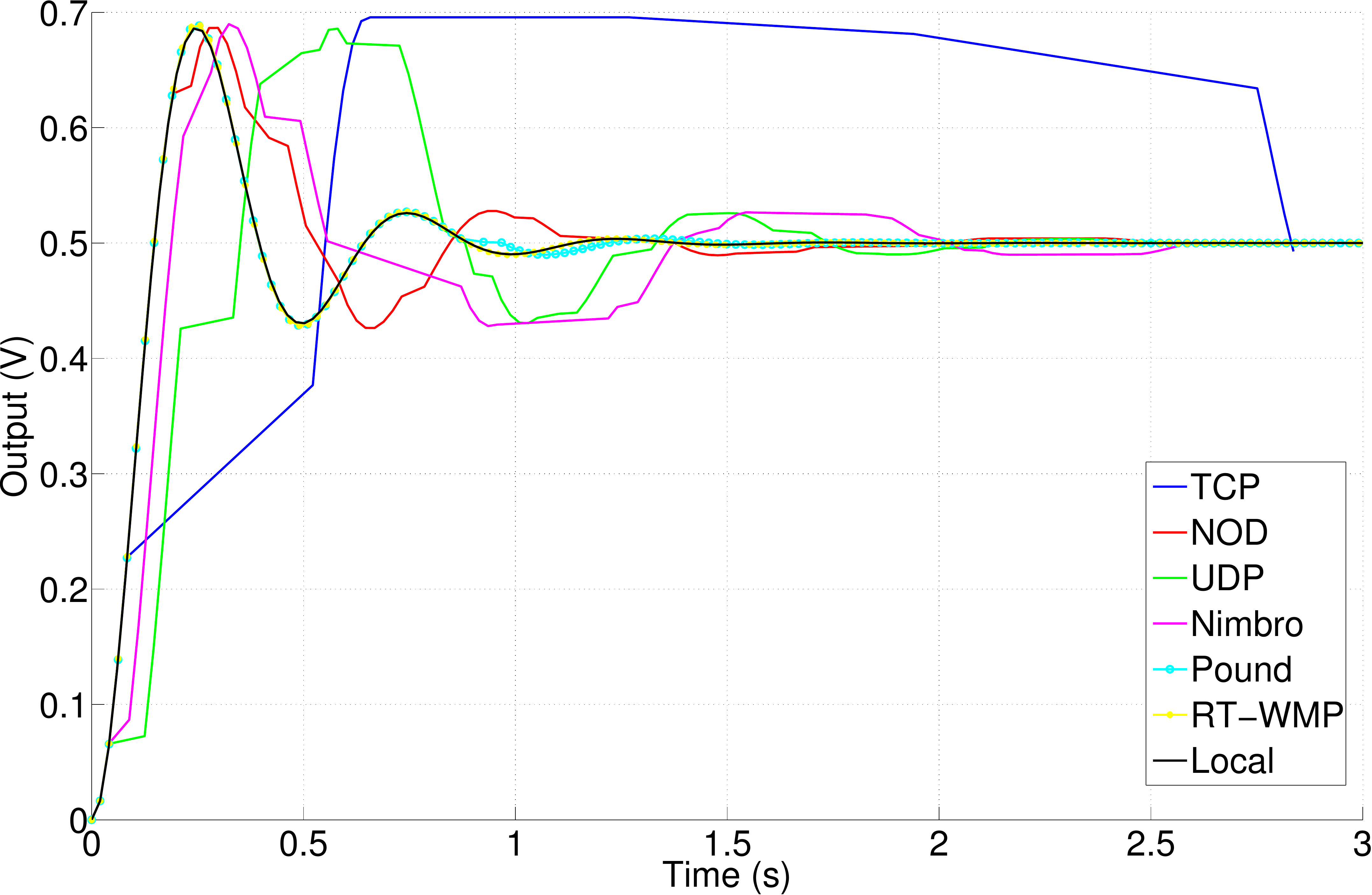}
\end{center}
\vspace{-3mm}

\caption{\label{fig:control}\small Output of the simple RLC system in a feedback control loop.}
\vspace{-3mm}
\end{figure}
The results in terms of the value of $Vc(k)$ over the time are shown in Figure \ref{fig:control}. Comparing them with the locally executed simulation, the \textit{Pound} and especially the \textit{RT-WMP} are the only methods capable of supporting control performance close to the local case while the other methods fail more or less noticeably due to jitter and  delay that the messages ---especially those of the $y$ flow that must compete with the perturbing flow--- suffer in their network trip.
It worth noting that the (this-time-small) differences between \textit{RT-WMP} and \textit{Pound} stem from the fact that \textit{RT-WMP} provides a system of \textit{global} priorities whereas \textit{Pound} only establish a priority between flows that originate in the same computer. This means that in case of \textit{Pound} the $u$ flow messages must compete with the perturbing flow's messages to access the medium, while with \textit{RT-WMP} their access are always given higher priority thanks to the protocol's MAC scheme.
Notice that when the experiment was conducted without perturbing flow, all the methods offered the same correct results, making the output indistinguishable from the locally executed. 

\subsection{More flows, more hops}
We also performed preliminary experiments with multiple flows in different directions and with multiple network nodes in a chain configuration that is, multiple hops.
The results confirm, as foreseeable, those in the configuration proposed in this paper.

On the one hand, additional hops cause a noticeable reduction of the available bandwidth. However, even when the required bandwidth is below the limit, the decreased end-to-end packet delivery ratio, the poor performance of TCP protocols in lossy networks and the congestion due to the lack of mechanisms for managing time-sensitive data or burst transmissions, make the delay and jitter increase noticeably at the same time the MDR decreases rapidly. 
On the other hand, the presence of multiple flows in the same direction worsen the situation not only due the increased required bandwidth but also because the flows interfere on each other as shown by the experiments.
In both scenarios, the \textit{Pound} offers solutions to mitigate this effects by introducing priorities and taking into account the time needed to send the packets over the network.

The introduction of flows in opposite direction (a forward flow from the base station to a teleoperated robot in the scenario proposed in this paper, for example) indirectly perturbs all the others since it competes with them to access the medium ---thus reduces the available bandwidth--- and can delay the delivery of more important or more priority messages.
However, since the access to the medium is resolved in a per-packet basis the influence is usually assumable. Even so, the use of a network-wide priority mechanism, as \textit{RT-WMP} does, improves the global behavior as shown in the experiment in section \ref{sec:control-loops} at the expense of additional overhead and thus results in a reduced throughput.

\section{Discussion}
\label{sec:discussion}
The experiments show that the standard ROS behavior, that by default uses the TCP protocol to connect topics of different nodes, is not completely adequate for wireless networks. The other options, like NOD and UDP offer better results, that however degrade when multiple flows compete to gain access to the medium. Mixed solutions, that is those in which UDP are used in conjunction with TCP flows, show a better behavior in terms of jitter and delay for UDP flows but degrade the latter flow that adapts to the bandwidth left by the former flow. This scheme could be taken into account for the scenarios considered in this paper, but similar issues to those shown with single-type transport would reappear if a system needs three of more flows in the same direction. 
Also, the experiments showed that the results sometimes depend on how the TCP (and NOD) flows adapt to the network and how they coexist with others.

On the other hand, the fact of having a single ROS core that acts as topics' name resolution system, increase the overhead, reducing the available bandwidth of the network besides limiting its flexibility.
The solutions based on multiple ROS cores such as \textit{Nimbro}, \textit{Pound} and \textit{RT-WMP} mitigate these problems but the results show that the former obtain similar results as the standard ROS solutions in terms of jitter, both for single-type (UDP/UDP) and mixed-type (UDP/TCP) solutions. This is due to the fact that one connection is created for each flow with the same problems in terms of coexistence and perturbation that they inflict on each other.
This suggests that using a unique connection for transporting all the different flows would improve the results and highlights the need for a system of priorities for topics with different requisites and importance.
This is the solution that both \textit{Pound} and \textit{RT-WMP} propose ---the former in a per-node  and the second in global terms--- showing better results especially in terms of jitter of the higher priority flow.

The improvement is even clearer in 3-nodes network where standard solutions show very poor results due to the congestion that the network suffers. Especially TCP-based methods, which has been designed to work in non-lossy networks, show huge message loss rates that makes their use impractical in systems in which a control loop is involved, both alone and in conjunction with UDP flows. This shows that using ROS over a multi-hop wireless link is likely to offer unsatisfactory results.
On the other hand, \textit{Nimbro} that performed reasonably well in 2-nodes network, suffers from UDP silent discards due to the congestion of the network either at the sender side or in the relay node. According to our experiments this is due to packets (in which large messages are split into) being queued in the Linux network layer in bursts, regardless of whether the queue has enough free space or not. As explained earlier, to avoid this, \textit{Pound} introduces a delay after any socket \textit{send()} call, that in principle leaves enough time for the packet to be sent over the network.

The results show that in this scenario \textit{Pound} outperforms all the other methods including \textit{RT-WMP} that pays a higher overhead and is incapable of maintaining a proper rate for the \textit{image} flow, as confirmed also by the bandwidth experiment results.

The control loop experiment confirms the results obtained in terms of jitter and delay showing that only the \textit{Pound} and \textit{RT-WMP} methods are capable of  properly closing the control loop and showing the desired response. The other methods, that comply with the control requisites when tested without the perturbing flow, fail more or less noticeably in the proposed scenario.

\section{Conclusions}
\label{sec:conclusions}
In this paper we presented a comparison of different solutions for using the Robotics Operating System (ROS) platform in wirelessly distributed systems.
Standard ROS communication protocols are in fact not optimized for wireless communication where aspects like reduced bandwidth (especially in multi-hop networks) and increased delay and jitter must be taken into account.

First, an analysis on ROS limitations in this aspect was presented, showing that the default choice for connecting topics can suffer from irregular delays and jitters (principally due to the Nagle's algorithm). Then, an analysis of the performance in presence of multiple flows (ROS topics) with different sizes and periods was performed showing that flows can interfere with each others timing making the implementation of tight control loops difficult.
Finally, we presented various multi-core solutions for ROS and proposed a new multi-core solution (called \textit{Pound}) that exploits flow priorities to meet the bandwidth and delay requirements of the system. We compared all the solutions including the standard ROS supported protocols with and without relay nodes in static wireless networks.

The results shows that flow prioritization is necessary if the reduction of  jitter is required, mainly in multi-hop networks. The paper aims to be a reference for those implementing wirelessly connected multi-robot and distributed systems, especially involving tight control loops. 
\balance

\section*{Acknowledgments}
This work has been funded by the projects ALERTA (CUD 2016-17) and ROBOCHALLENGE (DPI2016-76676-R) and programa Ibercaja-CAI de estancias de investigaci\'{o}n.

\bibliographystyle{IEEEtran}
\bibliography{references,WirelessRef_new}

\end{document}